\documentclass[aps,prb,onecolumn,groupedaddress,floatfix,longbibliography,superscriptaddress]{revtex4-2}

\usepackage{amsmath}
\usepackage{amssymb}
\usepackage{amsfonts}
\usepackage{bbold}
\usepackage{bm}
\usepackage{cancel}
\usepackage{times,float}
\usepackage{graphicx}
\usepackage[usenames,dvipsnames,svgnames]{xcolor}
\usepackage{hyperref}
\hypersetup{colorlinks=true, linkcolor=NavyBlue, citecolor=PineGreen,urlcolor=cyan}
\usepackage{multirow}
\usepackage{physics}
\usepackage{esint}
\usepackage{caption}
\usepackage{subcaption}
\usepackage{tikz}
\usepackage{pgfplots}
\usepackage{pgfplotstable}
\pgfplotsset{compat = newest}
\usetikzlibrary{positioning, arrows.meta}
\usepgfplotslibrary{fillbetween}
\usepackage{svg}

\begin{document}


\title{Landauer-based study of transport in Chern insulator heterostructures}

\author{J. Luna-Ramos}
\email{joseph.luna@unmsm.edu.pe}
\affiliation{Grupo de Física Teórica y Altas Energías, Universidad Nacional Mayor de San Marcos, Avenida Venezuela s/n Cercado de Lima, 15081, Lima, Perú}

\author{A. Mart\'{i}n-Ruiz}
\email{alberto.martin@nucleares.unam.mx}
\affiliation{Instituto de Ciencias Nucleares, Universidad Nacional Aut\'{o}noma de M\'{e}xico, 04510 Ciudad de M\'{e}xico, M\'{e}xico}

\begin{abstract}
We study charge transport through a trivial-topological-trivial junction described by the continuous Qi-Wu-Zhang model, which realizes a two-dimensional Chern-insulating phase. The central region is tuned into the topological regime, while the adjoining leads remain trivial, and an electrostatic barrier of tunable height and width is applied exclusively to the topological slab. By matching wave functions across the interfaces, we obtain the angle- and energy-resolved transmission probability and demonstrate the occurrence of Klein tunneling despite the presence of a bulk spectral gap. Within the continuum Dirac description, this perfect transmission originates from the inversion of the Dirac mass across the junction, which reflects the band inversion of the central layer relative to the trivial leads. In the Qi-Wu-Zhang model considered here, this mass inversion coincides with the transition between trivial and Chern-insulating phases and is accompanied by finite Berry curvature that governs the nonlinear transport response. The resulting transmission function is then incorporated into a Landauer-Büttiker framework to analyze both linear and nonlinear transport. Closed-form expressions for the linear and nonlinear conductances are derived at zero and finite temperatures. In addition, we investigate the role of dephasing, showing how partial loss of coherence suppresses Fabry-Pérot oscillations while leaving the overall transport trends intact. Finally, we map out the interplay between barrier height, slab thickness, and topological mass parameter, identifying optimal regimes that yield enhanced rectification in the nonlinear response.
\end{abstract}

\maketitle

\section{Introduction} \label{Intro}

The discovery of graphene opened a new era in condensed matter physics by demonstrating that materials can host quasiparticles mimicking relativistic Dirac fermions at accessible energy scales \cite{novoselov_two-dimensional_2005, zhang_experimental_2005, neto_electronic_2009}. Its linear dispersion near the Dirac points gives rise to remarkable transport properties such as high mobility, ambipolar conduction, and ballistic propagation over submicron distances \cite{bolotin_ultrahigh_2008, mayorov_dirac_2011}. Among the most striking manifestations of its Dirac-like character is Klein tunnelling: the counterintuitive prediction that massless relativistic fermions can transmit perfectly through high potential barriers at normal incidence. Originally introduced in particle physics as a solution of the Dirac equation in the presence of an energy barrier \cite{DOMBEY199941}, Klein tunnelling was brought into the condensed matter arena when Katsnelson, Novoselov and Geim identified it in graphene due to the chiral nature of its Dirac cones \cite{katsnelson_chiral_2006}. This effect can be understood within the Dirac approximation, where the low-energy spectrum around the Dirac points is described by an effective relativistic Hamiltonian. In graphene, this leads to a massless Dirac equation; however, there also exist systems where the relevant low-energy degrees of freedom are governed by a massive Dirac equation, allowing one to study the fate of Klein tunnelling and related relativistic phenomena in the presence of a bulk energy gap \cite{asboth_short_2016}. 
{Indeed, Klein tunnelling of massive Dirac fermions has been analyzed in a variety of condensed-matter settings, including graphene-based models with induced gaps \cite{Setare2010}, magnetically gapped topological-insulator junctions \cite{Goudarzi2015}, and three-dimensional systems hosting massive Kane fermions \cite{Betancur2017}, where finite or even perfect transmission arises from spinor matching, chirality conservation, or specific band-structure features.} {
These studies demonstrate that band inversion can strongly modify tunneling in massive Dirac systems.}

Beyond graphene, the focus has shifted toward engineered materials where Dirac or Weyl fermions acquire a mass and the system develops a nontrivial topology. In two dimensions, Chern insulators provide a paradigmatic example: their band inversion leads to a quantized Hall conductance without an external magnetic field, establishing a direct link between Berry curvature, topological invariants, and measurable transport properties \cite{haldane_model_1988, qi_topological_2006}. The ability to realize Chern phases in magnetically doped topological insulators has expanded the landscape of topological materials and created opportunities to study Klein tunnelling and related transport phenomena in systems where a bulk gap coexists with topological protection \cite{Wang2014,Lu2011}. {In this context, the Dirac mass acquires a qualitatively different meaning: rather than acting as a simple gap-opening parameter, it encodes band inversion and a nonzero Chern number, endowing the bands with a finite Berry curvature that actively constrains carrier dynamics.}

A natural setting to investigate these effects is through heterostructures that combine trivial and topological regions, where band inversion and interface scattering play a central role \cite{Liu2010, Linder2009}. At the interface between distinct topological phases, electron propagation is strongly influenced by the sign of the Dirac mass term and the associated Berry curvature \cite{xiao_berry_2010}. Unlike in graphene, where Klein tunnelling arises in a gapless spectrum \cite{katsnelson_chiral_2006}, in Chern insulators the phenomenon can survive despite the presence of a finite bulk gap, raising new questions about the interplay between barrier potentials, mass inversion, and transmission \cite{yang_topological_2015, huang_non-linear_2016}. {While related tunnelling studies in gapped systems have been reported previously \cite{Setare2010, Goudarzi2015, Betancur2017}, they do not address a trivial-topological-trivial heterostructure characterized by a well-defined Chern number, nor do they explore how Berry curvature and band inversion reshape the scattering problem beyond purely kinematic considerations.}

In this work we focus on a trivial-topological-trivial junction described by the continuous Qi-Wu-Zhang model \cite{qi_topological_2006}. The central region is tuned into the Chern-insulating regime, while the leads remain trivial, thereby defining an effective n-p-n junction. A gate-induced electrostatic barrier is applied to the topological slab, allowing us to study how potential profiles modulate transmission across a mass-inverted region. By solving the scattering problem explicitly, we demonstrate the existence of Klein tunnelling in this heterostructure and show how it is modified by the barrier height, slab thickness, and chemical potential \cite{katsnelson_chiral_2006, yang_topological_2015}. Our central result is that band inversion in the central region enables robust Klein-like transmission through a gapped slab embedded between insulating leads. Within the Qi-Wu-Zhang Chern-insulator model, this inversion of the Dirac mass corresponds to the transition between trivial and Chern phases,  providing a direct connection between topological band inversion and the scattering properties of bulk states. Furthermore, we show that Berry curvature plays a central role in nonlinear transport, where it enters explicitly in the Landauer formulation  of the nonlinear Hall conductance.

{
Building on the scattering analysis discussed above, we next formulate the transport problem within a Landauer-B\"uttiker framework \cite{landauer_electrical_1957, buttiker_four-terminal_1986} to evaluate both longitudinal and transverse conductances at linear and nonlinear order. } Numerical calculations are complemented with analytic approximations valid in the limits of large chemical potential and near the band edge. These results not only provide a detailed map of transport regimes but also connect directly to realistic material platforms such as Cr-doped (Bi,Sb)$_2$Te$_3$, where Chern-insulating phases have been realized experimentally \cite{chang_experimental_2013}.  { A key novelty of our analysis is the systematic treatment of linear and nonlinear transport on equal footing: we show that higher-order longitudinal conductance coefficients are strongly enhanced near the band edges of the Chern region, and we demonstrate that the nonlinear Hall conductance, controlled by Berry curvature and finite transmission, vanishes inside the bulk gap and emerges only once the chemical potential enters the bands.
}

By bridging the relativistic transport phenomena first observed in graphene \cite{novoselov_two-dimensional_2005, katsnelson_chiral_2006} with the topological responses of Chern insulators \cite{haldane_model_1988, yu_quantized_2010}, our study contributes to a broader understanding of how topology, Berry curvature, and Klein tunnelling conspire to shape both linear and nonlinear transport in realistic topological heterostructures. 
{This unified perspective, combining topology-driven scattering with nonlinear transport in a realistic junction geometry, provides the main motivation for the extensive model study presented in this work.}

The remainder of this paper is organized as follows. In Sec. \ref{QWZ_model} we introduce the Qi-Wu-Zhang model in the continuum limit and develop its Dirac description, highlighting the essential ingredients of a Chern insulator and discussing its basic spectral and topological properties. Section \ref{Klein_tunneling_section} is devoted to the study of electron transmission through a gate-induced barrier that defines a trivial-topological-trivial junction; {here we demonstrate the emergence of Klein tunnelling as a consequence of band inversion and the associated spinor matching across the junction.} In Sec. \ref{longitudinal_conductance_section} we analyze the longitudinal conductance within the Landauer framework, presenting both linear and nonlinear contributions together with analytic asymptotics and numerical results. Section \ref{Hall_conductance_section} extends this analysis to the transverse (Hall) conductance, again treating linear and nonlinear responses on equal footing and identifying the regimes where nonlinear Hall effects are most pronounced. Finally, Sec. \ref{Conclusions} summarizes our main conclusions and discusses possible extensions of this work.

\section{Qi-Wu-Zhang model in the continuum limit: Dirac description of a Chern insulator} \label{QWZ_model}

We consider the Qi-Wu-Zhang model as a minimal description of a Chern insulating phase in two dimensions \cite{qi_topological_2006}. The starting point is a tight-binding Hamiltonian defined on a square lattice with lattice spacing $a$, where each site hosts a two-component spinor degree of freedom. In real space, the Hamiltonian is given by
\begin{align}
    H = \sum _{\mathbf{r}} \left[ t _{x} \hat{c} _{\mathbf{r}} ^{\dagger} \sigma _{x}  \hat{c} _{\mathbf{r} + \hat{x}} + t _{y} \hat{c} _{\mathbf{r}} ^{\dagger} \sigma _{y}  \hat{c} _{\mathbf{r} + \hat{y}} + m _{0}  \hat{c} _{\mathbf{r}} ^{\dagger} \sigma _{z}  \hat{c} _{\mathbf{r}} - B \sum _{\boldsymbol{\delta} = \pm a \hat{x} , \pm a \hat{y} } \hat{c} _{\mathbf{r}} ^{\dagger} \sigma _{z}  \hat{c} _{\mathbf{r} + \boldsymbol{\delta} } \right] + \mbox{H.c.} , 
\end{align}
where $\hat{c} _{\mathbf{r}} ^{\dagger} = (\hat{c} _{\mathbf{r} \uparrow} ^{\dagger},\hat{c} _{\mathbf{r} \downarrow} ^{\dagger})$ is a spinor creation operator at lattice site $\mathbf{r}$, and $\sigma _{x,y,z}$ are the Pauli matrices acting in the internal pseudospin space.  The hopping amplitudes $t_{x}$  and  $t_{y}$ mediate spinor-flipping nearest-neighbor couplings along the $x$ and $y$ directions through the $\sigma _{x}$ and $\sigma _{y}$ channels, while the onsite term $m _{0} \sigma _{z}$ breaks time-reversal symmetry and opens a gap. The last term, proportional to $B$, corresponds to spin-conserving hopping in the $\sigma _{z}$ channel between nearest neighbors. This extension is crucial for inducing a momentum-dependent mass term and ensuring a well-defined topological phase.

To analyze the bulk band structure, we perform a Fourier transform to momentum space using
\begin{align}
    \hat{c} _{\mathbf{r}} = \sum _{\mathbf{k}} e ^{i \mathbf{k} \cdot \mathbf{r}} \, \hat{c} _{\mathbf{k}} , 
\end{align}
where the crystal momentum $\mathbf{k}$ lives in the Brillouin zone $\mathbf{k} \in [-\pi / a , \pi / a]$. The Hamiltonian becomes diagonal in $\mathbf{k}$, taking the form
\begin{align}
    H = \sum _{\mathbf{k}}  \hat{c} _{\mathbf{k}} ^{\dagger} \left[ \boldsymbol{\sigma} \cdot \mathbf{h} (\mathbf{k}) \right] \hat{c} _{\mathbf{k}} , 
\end{align}
with
\begin{align}
    \mathbf{h} (\mathbf{k}) = (t _{x} \sin ( k _{x}a) ,t _{y} \sin (k _{y} a ), m _{0} - 2B \left[ \cos (k _{x}a) + \cos ( k _{y} a) \right] ) . 
\end{align}
In this expression, the hopping in the $\sigma _{z}$ channel generates the cosine terms in the $h _{z} (\mathbf{k}) $ component. To facilitate the analysis of the low-energy limit, we rewrite this component as 
\begin{align}
    h _{z} (\mathbf{k}) = m_{0} -4B + 2B \left[  1 - \cos ( k _{x} a) \right]  + 2B \left[  1 - \cos ( k _{y} a) \right]  ,
\end{align}
so that $h _{z} (\mathbf{0}) = m _{0} -4B$, identifying clearly the momentum-independent shift in the effective mass due to the $B$-term.

To obtain a continuum description valid at low energies, we expand the Hamiltonian near the $\Gamma$ point, where $\vert \mathbf{k} \vert \ll \pi /a$. Using the approximations $\sin (k _{i} a) \approx k _{i} a$ and $\cos ( k _{i} a) \approx 1 - \frac{1}{2} (k _{i} a) ^{2}$ for $i=x,y$, we obtain
\begin{align}
    \mathbf{h} (\mathbf{k}) \approx (t _{x} a  k _{x},t _{y} a k _{y}, m _{\mbox{\scriptsize eff}} + B a ^{2} k ^{2} ) ,
\end{align}
where $m _{\mbox{\scriptsize eff}} = m _{0} - 4B$ and $k ^{2} = k _{x} ^{2} + k _{y} ^{2}$. This yields the effective low-energy Hamiltonian in the continuum limit:
\begin{align}
    H _{\mbox{\scriptsize Dirac}} (\mathbf{k}) = \hbar v _{x} k _{x} \sigma _{x} + \hbar v _{y} k _{y} \sigma _{y} + ( m _{\mbox{\scriptsize eff}} + \beta \hbar ^{2} k ^{2}) \sigma _{z} , \label{Dirac_Hamiltonian}
\end{align}
where $v _{i} = at_{i}/ \hbar$ is the Fermi velocity along the $i$ direction and the coefficient $\beta = Ba ^{2} / \hbar ^{2}$ controls the quadratic correction to the mass. The linear terms describe a massive Dirac fermion in two dimensions, while the $k ^{2}$ term regularizes the dispersion at higher momenta.

The topology of the system is determined by the sign of the mass term at $\mathbf{k} = \mathbf{0}$, as it controls the global winding of the eigenstates in momentum space \cite{xiao_berry_2010}. For the effective Hamiltonian (\ref{Dirac_Hamiltonian}) the eigenvalues are $E _{s} (\mathbf{k}) = s \, \vert \mathbf{d} (\mathbf{k}) \vert $, where $\mathbf{d} (\mathbf{k})$ is the vector appearing if we decompose the Hamiltonian as $H _{\mbox{\scriptsize Dirac}} (\mathbf{k}) =  \boldsymbol{\sigma} \cdot \mathbf{d} (\mathbf{k})$ and $s= \pm 1$ is the band index. The normalized eigenstates of the conduction ($+$) and valence ($-$) bands can be written as
\begin{align}
    \ket{u _{s} (\mathbf{k})} = \frac{1}{\sqrt{2}} \left( \begin{array}{c}
         \sqrt{1 + s \cos \theta _{\mathbf{k}} }  \\[4pt]
          s \sqrt{1 - s \cos \theta _{\mathbf{k}} } \, e ^{i \phi _{\mathbf{k}} } 
    \end{array} \right)  ,
\end{align}
where $\theta _{\mathbf{k}} = \arccos (d _{z} / \vert \mathbf{d} \vert )$ and $\phi _{\mathbf{k}} = \arg (d _{x} + i d _{y})$ are the spherical angles defining the orientation of the vector $\mathbf{d} (\mathbf{k})$, i.e.,
\begin{align}
    \hat{\mathbf{d}} (\mathbf{k}) = \frac{\mathbf{d} (\mathbf{k}) }{\vert \mathbf{d} (\mathbf{k}) \vert } = (\sin \theta _{\mathbf{k}} \cos \phi _{\mathbf{k}} , \sin \theta _{\mathbf{k}} \sin \phi _{\mathbf{k}} , \cos \theta _{\mathbf{k}}) .
\end{align}
The Chern number of the lower band is given by the integral of the Berry curvature over the entire Brillouin zone. In the continuum limit, this reduces to the well-known formula \cite{asboth_short_2016}
\begin{align}
    C = \frac{1}{4 \pi } \int _{\mbox{\scriptsize BZ}} d ^{2} \mathbf{k} \;  \hat{\mathbf{d}} \cdot (\partial _{k _{x}} \hat{\mathbf{d}} \times \partial _{k _{y}} \hat{\mathbf{d}} ) ,
\end{align}
which counts the number of times the unit vector $\hat{\mathbf{d}} (\mathbf{k})$ wraps the Bloch sphere as $\mathbf{k}$ spans the Brillouin zone. This topological invariant is quantized and robust against perturbations that do not close the energy gap.

In our model, the vector $\hat{\mathbf{d}} (\mathbf{k})$ asymptotically points along the $+z$-direction as $\vert \mathbf{k} \vert \to \infty$, due to the quadratic mass term $\beta \hbar ^{2} k ^{2}$. At $\mathbf{k} = \mathbf{0}$, the mass term equals $m _{\mbox{\scriptsize eff}}$. If $m _{\mbox{\scriptsize eff}} <0$, then $\hat{\mathbf{d}} (\mathbf{k})$ interpolates between the south and north poles of the Bloch sphere, covering it exactly once. This yields $C= \pm 1$, with the sign depending on the orientation of $\hat{\mathbf{d}} (\mathbf{k})$. If $m _{\mbox{\scriptsize eff}}>0$, the vector field $\hat{\mathbf{d}} (\mathbf{k})$ points toward the north pole everywhere, and the map is topologically trivial, giving $C= 0$. The topological phase transition occurs at $m _{\mbox{\scriptsize eff}} = 0$, where the bulk gap closes at $\mathbf{k} = \mathbf{0}$.

\begin{figure}
    \centering
    \includegraphics[width=0.48\linewidth]{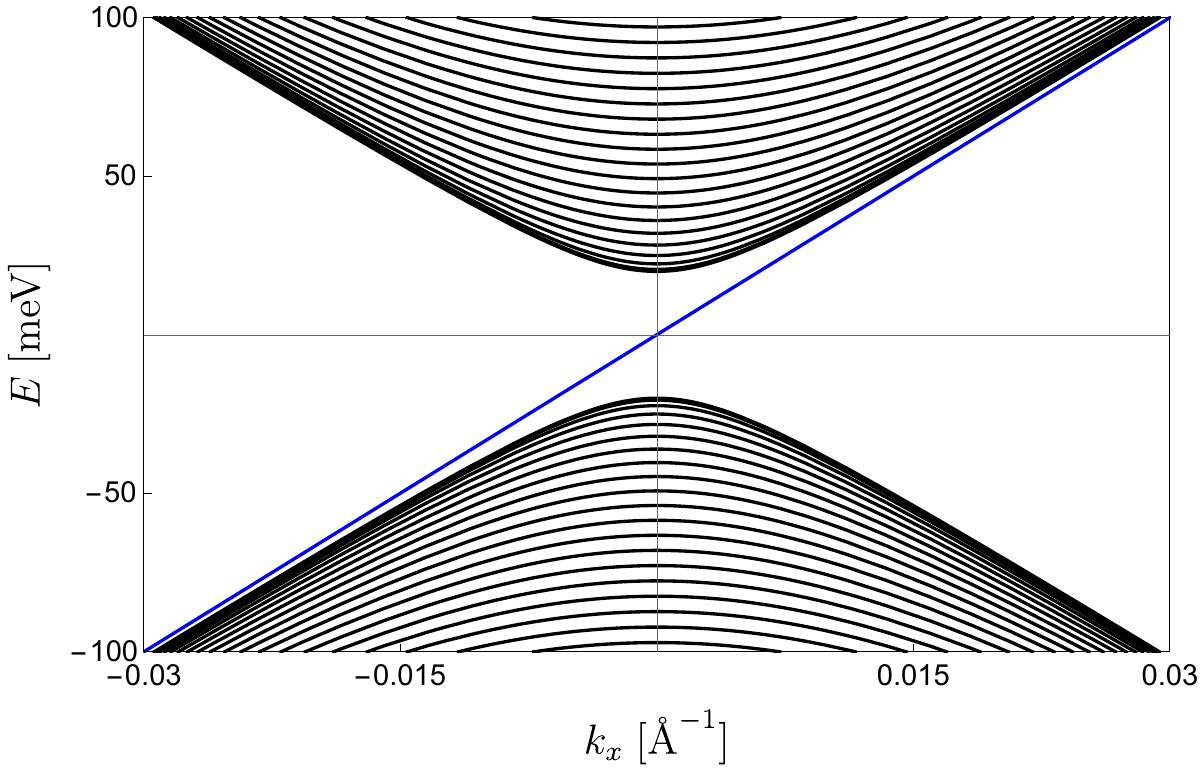} \hspace{1cm}
    \begin{tikzpicture}[]
        \begin{axis}[
            scale = 1,
            xmin = -2.2, xmax = 2.2,
            ymin = -0.5, ymax = 1.7,
            width=8cm, height=6cm,
            axis lines* = center,
            xtick = {0}, ytick = \empty,
            clip = false,
            axis line style={latex-latex},
            ytick={1},
            yticklabel style={yshift=-1.7ex, anchor=west},
            extra y ticks={0},
            extra y tick style={yticklabel style={yshift=-1.7ex, anchor=east}}]
            \node [right=10pt,font=\small] at (current axis.right of origin) {$m _{\mbox{\scriptsize eff}}$};
            \node [above=10pt,font=\small] at (current axis.above origin) {$C_s$};
            
            \addplot[blue,domain=-1.9:0] {1};
            \addplot[blue,domain=0:1.9] {0};
            \addplot[blue,mark=*,fill=white] coordinates {(0,0)};
            \addplot[blue,mark=*,fill=white] coordinates {(0,1)};
        \end{axis}
        \end{tikzpicture}
    \caption{Left: Band structure of the Chern insulator, showing the linear dispersion of the edge state crossing the bulk gap. Right: Chern number as a function of the effective mass $m _{\mbox{\scriptsize eff}}$, illustrating that the sign inversion of the mass is accompanied by a change in the topological invariant.}
    \label{fig:placeholder}
\end{figure}
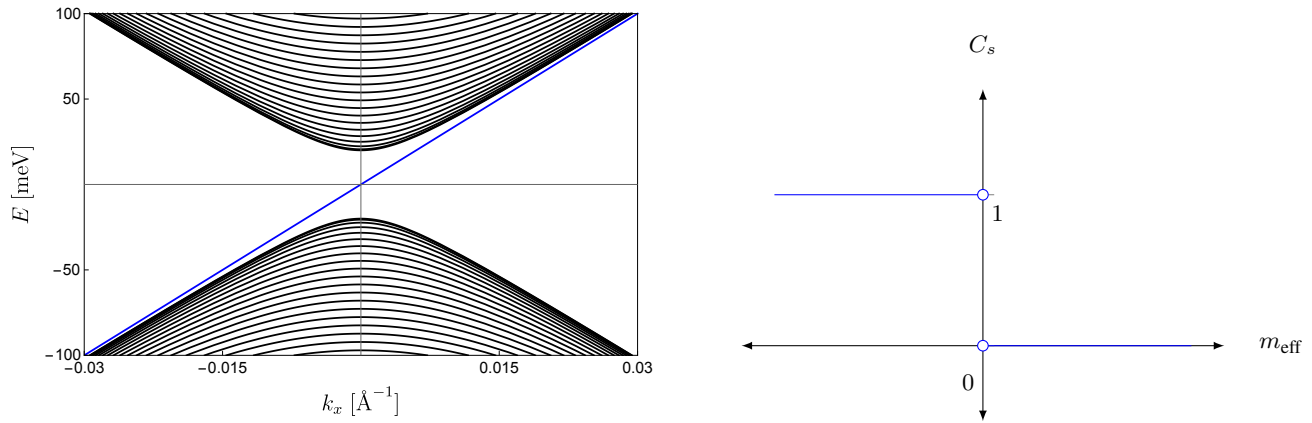

This continuum Dirac formulation, grounded in the lattice model with explicit lattice spacing $a$, provides a transparent and physically consistent framework for studying low-energy transport phenomena. In particular, it allows the introduction of smooth, spatially varying mass profiles $m _{\mbox{\scriptsize eff}} (\mathbf{r})$ and external electrostatic potentials. {Before turning to the scattering analysis, we briefly comment on an experimentally relevant implementation of the mass-inverted junction considered below and on the realistic parameter scales.}

{

A realistic material platform for the junction considered in this work is provided by magnetic topological insulators based on (Bi,Sb)$_2$Te$_3$. The central region can be realized using Cr-doped (Bi,Sb)$_2$Te$_3$, which breaks time-reversal symmetry and exhibits a quantized anomalous Hall effect with Chern number $C = \pm 1$ \cite{yu_quantized_2010, chang_experimental_2013}, while the outer regions, composed of the undoped compound, remain topologically trivial ($C=0$). This configuration enables the engineering of a mass-inverted profile $m(x)$ across the interfaces while maintaining structural and electronic compatibility. Although the model employed is formulated on a square lattice, as in the prototypical Qi-Wu-Zhang Hamiltonian \cite{qi_topological_2006}, it effectively captures the essential low-energy features of the system near the $\Gamma$ point. Despite the fact that real materials crystallize in a layered rhombohedral structure with hexagonal symmetry \cite{zhang_topological_2009}, their band structure exhibits Dirac-like behavior that justifies the use of such simplified lattice models. This approach is widely used to study bulk topological and transport phenomena in Chern insulators \cite{asboth_short_2016, hasan_colloquium_2010}.

Within this minimal continuum junction description, we take the magnitude of the Dirac mass $|m _{\mbox{\scriptsize eff}}|$ to be uniform and implement the trivial-topological-trivial heterostructure through a sign reversal of $m _{\mbox{\scriptsize eff}} (x)$ in the central region, which captures the band inversion while keeping the bulk gap $2|m _{\mbox{\scriptsize eff}}|$ fixed. This sign change can be physically realized via magnetic doping, strain, or thickness control \cite{Liu2010, yu_quantized_2010, chang_experimental_2013}. Realistic parameters include Fermi velocities $v_F \sim 4\text{ - }6 \times 10^{5}$~m/s and a Dirac mass $|m _{\mbox{\scriptsize eff}}| \sim 20$ - $30$~meV, corresponding to a bulk gap of $2|m _{\mbox{\scriptsize eff}}| \sim 40$ - $60$~meV \cite{zhang_topological_2009, xia_observation_2009}. For concreteness, and unless stated otherwise, we adopt representative values within this range in the calculations presented below. The resulting band structure for this parameter regime, together with the associated Chern number as a function of the effective mass and the corresponding topological phase transition, is illustrated in Fig.~\ref{fig:placeholder}. It is important to emphasize that, within the QWZ model, the sign of the effective Dirac mass is not merely a gap parameter but directly determines the topological invariant of the bulk bands. 
As discussed above, the change of sign of $m_{\mathrm{eff}}$ corresponds to a band inversion that drives the transition between the trivial phase ($C=0$) and the Chern-insulating phase ($C=\pm 1$). 
Therefore, the mass inversion implemented in the junction is the low-energy manifestation of the topological phase transition of the underlying lattice model. 
In the scattering formulation developed below, this inversion appears as a sign change of the Dirac mass entering the continuum Hamiltonian, which governs the matching of spinor wavefunctions across the interfaces.

}

{With this physical picture in mind, we now employ the continuum Dirac formalism to model a finite topological region embedded in a trivial background and to compute the transmission of bulk states across the resulting junction. We then use the resulting transmission function to evaluate linear and nonlinear conductances within the Landauer framework at zero temperature \cite{datta_electronic_1995}.}

\section{Klein tunneling in a topological junction} \label{Klein_tunneling_section}

To investigate the transport properties of the system, we consider the low-energy regime where the quadratic momentum correction to the mass can be safely neglected. Furthermore, we restrict ourselves to the isotropic case where the effective Fermi velocity is the same along both spatial directions, i.e., $v _{x} = v _{y} = v _{F}$. Under these assumptions, and in the presence of an external electrostatic potential $V(\mathbf{r})$ and a mass profile $m(\mathbf{r})$, the effective Hamiltonian (\ref{Dirac_Hamiltonian}) reduces to a simplified Dirac form given by
\begin{align}
    H _{\mbox{\scriptsize Dirac}} (\mathbf{k}) = \hbar v _{F} ( k _{x} \sigma _{x} + k _{y} \sigma _{y} ) +  m (\mathbf{r}) \, \sigma _{z} + V(\mathbf{r}) \, \sigma _{0} , \label{Dirac_Hamiltonian_linear}
\end{align}
where $\mathbf{k} = (k _{x},k_{y})$ is the crystal momentum, $\mathbf{r} = (x,y)$ is the position and $m$ is the Dirac mass. For notational simplicity, we omit the subscript eff and simply write $m$ for the effective mass parameter. The sign of $m (\mathbf{r})$ determines the local topological character of each region: a negative mass corresponds to a Chern insulating phase, while a positive mass indicates a topologically trivial insulator \cite{haldane_model_1988}. Within the continuum Dirac formulation, the transport problem can be solved independently of the explicit value of the topological invariant, and the transmission properties are governed by the matching of spinor wavefunctions across regions with different mass parameters.  In the present QWZ-based heterostructure, however, the sign reversal of the Dirac mass is not an arbitrary parameter change but reflects the band inversion associated with the transition between trivial and Chern-insulating phases. 
Thus, the mass-inverted junction considered here provides a concrete realization of a trivial-topological-trivial interface in which the scattering problem directly probes the consequences of topological band inversion.

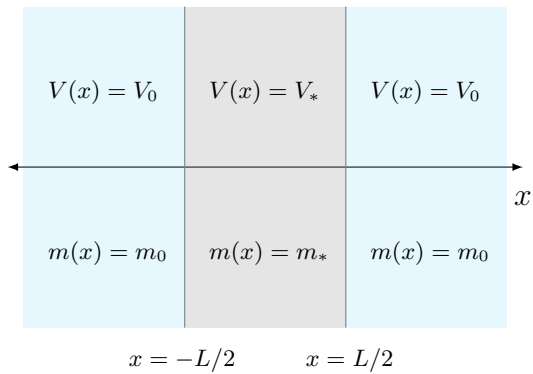
\begin{figure}
    \centering
    \begin{tikzpicture}
        \begin{axis}[
            scale = 1.0,
            xmin = -3.2, xmax = 3.2,
            ymin = -2.0, ymax = 2.0,
            axis lines* = center,
            axis y line=none,
            ytick = \empty,
            xtick = \empty,
            clip = false,
            axis line style={latex-latex},
            ]
            
            \addplot +[gray,mark=none] coordinates {(1, -1.5) (1, 1.5)};
            \addplot +[gray,mark=none] coordinates {(-1, -1.5) (-1, 1.5)};
            \fill[gray, opacity = 0.2] (1, -1.5) -- (1, 1.5) -- (-1, 1.5) -- (-1, -1.5);
            \fill[cyan, opacity = 0.1] (1, -1.5) -- (1, 1.5) -- (3, 1.5) -- (3, -1.5);
            \fill[cyan, opacity = 0.1] (-3.0, -1.5) -- (-3.0, 1.5) -- (-1.0, 1.5) -- (-1.0, -1.5);
            \node [right=10pt,below=6pt,font=\large] at (current axis.right of origin) {$x$};
            \node [right,font=\small] at (-0.8, 0.7) {$V(x)=V_{\ast}$};
            \node [right,font=\small] at (-2.8, 0.7) {$V(x)=V_0$};
            \node [right,font=\small] at (1.2, 0.7) {$V(x)=V_0$};
            \node [right,font=\small] at (-0.8, -0.8) {$m(x)=m_{\ast}$};
            \node [right,font=\small] at (-2.8, -0.8) {$m(x)=m_0$};
            \node [right,font=\small] at (1.2, -0.8) {$m(x)=m_0$};
            \node [right,font=\small] at (0.4, -1.8) {$x=L/2$};
            \node [right,font=\small] at (-1.8, -1.8) {$x=-L/2$};
        \end{axis}
    \end{tikzpicture}
    \caption{Schematic representation of the trivial-topological-trivial heterostructure. The central region (shown in gray), characterized by mass parameter $m_{\ast}$ and potential $V_{\ast}$, realizes the topological insulating phase, while the outer regions (shown in cyan) with mass $m_{0}$ and potential $V_{0}$ remain in the trivial regime. The configuration forms a ribbon-like geometry that allows for the study of transport across topological and trivial domains.}
    \label{fig:energypotencial}
\end{figure}

In the following analysis, we consider spatial profiles $m(x)$ and $V(x)$ that vary only along the $x$-direction, as shown in Fig. \ref{fig:energypotencial}. The transmission of bulk Dirac carriers through such a junction can then be computed by solving the Dirac equation in each region and matching the spinor wavefunctions at the interfaces. We consider a mass profile $m(x)$ that is piecewise constant, consisting of three distinct regions. The outer regions ($\vert x \vert > L/2$) are assumed to be topologically trivial, with a positive and uniform mass parameter $m_{0}>0$. In contrast, the central region ($\vert x \vert < L/2$) is characterized by a tunable mass $m_{\ast}$, whose sign determines the topological character of the interface. When $m_{\ast} <0$, the central region corresponds to a Chern insulating phase, forming a nontrivial junction between trivial insulating leads. If $m_{\ast}>0$, the entire system remains topologically trivial. Thus, the mass profile is defined as
\begin{align}
    m(x) = \left\lbrace \begin{array}{c}
         m _{\ast} , \\[7pt]
         m _{0} ,  
    \end{array} \right. \quad \begin{array}{c}
         \mbox{for } \vert x \vert < L/2 , \\[7pt]
        \mbox{for } \vert x \vert > L/2  
    \end{array} .
\end{align}
We also consider an electrostatic potential profile of the form
\begin{align}
    V(x) = \left\lbrace \begin{array}{c}
         V _{\ast} , \\[7pt]
         V _{0} ,  
    \end{array} \right. \quad \begin{array}{c}
         \mbox{for } \vert x \vert < L/2 , \\[7pt]
        \mbox{for } \vert x \vert > L/2  
    \end{array} ,
\end{align}
where $V _{\ast}$ represents a gate-induced potential barrier in the central region and $V_{0}$ denotes the electrochemical potential of both the left and right leads.  The central potential $V _{\ast}$ can be tuned by applying a gate voltage locally over the topological region, thereby modifying the energy landscape experienced by the carriers \cite{xia_observation_2009, cho_gate-tunable_2011}. Meanwhile, the value of $V_{0}$ determines the chemical potentials of the reservoirs and controls the injection and extraction of electrons through the corresponding Fermi distribution functions \cite{datta_electronic_1995, buttiker_four-terminal_1986}. These external biases play a crucial role in driving current through the system and allow one to explore different transport regimes \cite{sodemann_quantum_2015}. Importantly, the electrostatic potential affects only the scalar part of the Hamiltonian and does not alter the topological character of the system, which remains governed by the sign of the mass term \cite{haldane_model_1988, qi_topological_2006}. A schematic representation of the spatial profiles of the mass and electrostatic potential is shown in Fig. \ref{fig:energypotencial}. 

To compute the transmission probability through the heterostructure, we solve the stationary Dirac equation at a fixed energy $E$. Since both the mass term $m(x)$ and the electrostatic potential $V(x)$ are piecewise constant, the solutions in each region can be written as plane-wave spinors corresponding to the same energy eigenvalue, but with opposite directions of propagation along the transport axis. That is, within each region, the general solution is a linear combination of right- and left-moving eigenstates with energy $E$, corresponding to incident, reflected, and transmitted components of the same quasiparticle branch.

In the region $\vert x \vert > L/2$, the energy of the fermion takes the value $E = V _{0} + s \sqrt{ \hbar ^{2} v _{F} ^{2} k ^{2}  + m _{0} ^{2} }$, where $s = \mbox{sgn} (E -V _{0})$. The plane-wave Dirac spinors are $\psi _{s} ^{(\xi)} (x,y) = \phi _{s} ^{(\xi)} (x) \, e ^{i ( k _{y} y - Et ) } $, being $\phi _{s} ^{(\xi)} (x)$ a two-component column vector of the form
\begin{align}
    \phi _{s} ^{(\xi)} (x) =  \sqrt{\frac{ E  - V _{0} + m _{0} }{ 2 ( E - V _{0} ) }}  \left( \begin{array}{c} 1  \\[7pt]      \frac{ \hbar v _{F} k }{ E - V _{0} + m _{0} } \, e ^{i \theta _{\xi} (\mathbf{k}) } 
    \end{array} \right) \, e ^{i \xi k _{x} x}  ,
\end{align}
where $\theta _{\xi} (\mathbf{k}) = \arctan ( \xi k _{y}/k_{x})$, $k ^{2} = k _{x} ^{2} + k _{y} ^{2} $ and $k_{x} = \sqrt{ \frac{(E - V _{0}) ^{2} - m _{0} ^{2}}{\hbar ^{2} v _{F} ^{2} } - k _{y} ^{2} }$. Here, the index $\xi = \pm$ labels the direction of propagation along the $x$-axis: $\xi = + $ corresponds to right-moving states (with positive group velocity), while $\xi = - $ denotes left-moving states. This notation will be used uniformly across all regions to distinguish between forward- and backward-propagating components of the wavefunction at a given energy.

In the central region $\vert x \vert < L/2$, the energy of the carriers takes the value $E = V _{\ast} + s _{\ast} \sqrt{ \hbar ^{2} v _{F} ^{2} k _{\ast} ^{2}  + m _{\ast} ^{2} }$, where $s _{\ast} = \mbox{sgn} (E -V _{\ast})$. The corresponding plane-wave Dirac spinors can be written as $\tilde{\psi} _{s _{\ast}} ^{(\xi)} (x,y) = \tilde{\phi} _{s _{\ast}} ^{(\xi)} (x) \, e ^{i ( k _{y} y - Et ) } $, where
\begin{align}
    \tilde{\phi} _{s _{\ast}} ^{(\xi)} (x) =  \sqrt{\frac{ E  - V _{\ast} + m _{\ast} }{ 2 ( E - V _{\ast} ) }}  \left( \begin{array}{c} 1  \\[7pt]      \frac{ \hbar v _{F} k _{\ast} }{ E - V _{\ast} + m _{\ast} } \, e ^{i \varphi _{\xi} (\mathbf{k}) }     \end{array} \right) \, e ^{i \xi q _{x} x}  ,
\end{align}
where $k _{\ast} ^{2} = q _{x} ^{2} + k _{y} ^{2}$, $\varphi _{\xi} (\mathbf{k} _{\ast}) = \arctan ( \xi k _{y}/q _{x})$ and $q _{x} = \sqrt{ \frac{(E - V _{\ast}) ^{2} - m _{\ast} ^{2}}{\hbar ^{2} v _{F} ^{2} } - k _{y} ^{2} }$. Once the spinor eigenstates are known, the general solution in each region is constructed as a linear combination of these states, weighted by reflection, transmission, or matching coefficients. In the left region ($ x < L/2$), the wavefunction includes an incident and a reflected component:
\begin{align}
    \Psi _{L} (x,y) = \psi _{s} ^{(+)} (x,y) + \mathcal{R} \, \psi _{s} ^{(-)} (x,y) , 
\end{align}
where $\mathcal{R}$ is the reflection amplitude. In the central region ($\vert x \vert < L/2$), the wavefunction is a superposition of right- and left-moving states:
\begin{align}
    \Psi _{C} (x,y) = \mathcal{A} \, \tilde{\psi} _{s _{\ast}} ^{(+)} (x,y) + \mathcal{B} \, \tilde{\psi} _{s _{\ast}} ^{(-)} (x,y) , 
\end{align}
with coefficients $\mathcal{A}$ and $\mathcal{B}$ determined by the boundary conditions. In the right region ($ x > L/2$), only a transmitted wave is present:
\begin{align}
    \Psi _{R} (x,y) = \mathcal{T} \, \psi _{s} ^{(+)} (x,y) ,
\end{align}
where $\mathcal{T}$ is the transmission amplitude. Continuity of the wavefunction at $x = \pm L/2$ imposes matching conditions on the spinor components, leading to a system of linear equations for the coefficients $\mathcal{A}$, $\mathcal{B}$, $\mathcal{R}$ and $\mathcal{T}$. Solving this system yields the transmission probability (defined by $T = \vert \mathcal{T} \vert $): 
\begin{align}
    T (E,k_{y}) = \frac{4 \Gamma _{ss_{\ast}} ^{2} \cos ^{2} \theta _{\mathbf{k}} \cos ^{2} \varphi _{\mathbf{k}} }{4 \Gamma _{ss_{\ast}} ^{2} \cos ^{2} \theta _{\mathbf{k}} \cos ^{2} \varphi _{\mathbf{k}} \cos ^{2} (q _{x} L) + \sin ^{2} (q _{x} L) \left( 1 + \Gamma _{ss_{\ast}} ^{2} - 2 \Gamma _{ss_{\ast}} \sin \theta _{\mathbf{k}} \sin \varphi _{\mathbf{k}} \right) ^{2}  } , \label{Transmission1}
\end{align}
where $\theta _{\mathbf{k}} = \theta _{+} (\mathbf{k})$, $\varphi _{\mathbf{k}} = \varphi _{+} (\mathbf{k})$ and 
\begin{align}
    \Gamma _{ss_{\ast}} = \frac{ (E - V _{0} + m _{0}) ( \hbar v _{F}  k _{\ast} ) }{ (E - V _{\ast} + m _{\ast}) ( \hbar v _{F} k ) } .
\end{align}
To facilitate the analysis and visualization of the transmission probability, it is convenient to express all quantities in dimensionless form. This allows for a clearer interpretation of the relevant parameters and a direct comparison across different regimes. In the following we consider $V_{0}=0$ and introduce the following dimensionless parameters: $\epsilon = V _{\ast} / E$, $\lambda _{0} = m _{0} / E$, $\lambda _{\ast} = m _{\ast} / E$ and $\kappa = EL/\hbar v _{F}$. On the other hand, the incident  and transmitted angles are not independent and must be properly related. The conservation of transverse momentum $k _{y}$ across the interfaces implies a constraint between the longitudinal wavevectors in each region, which can be naturally interpreted as a generalized Snell's law for Dirac fermions. That is, the angles of incidence and transmission, defined via the relations $\theta _{\mathbf{k}} = \arctan ( k _{y}/k_{x})$ and $\varphi _{\mathbf{k}} = \arctan ( k _{y}/q _{x})$, are connected through
\begin{align}
    k \sin \theta _{\mathbf{k}} = k _{\ast} \sin \varphi _{\mathbf{k}} ,
\end{align}
where $k = \frac{1}{\hbar v _{F}} \sqrt{  E ^{2} - m _{0} ^{2} }$ and $k _{\ast} = \frac{1}{\hbar v _{F}} \sqrt{(E - V _{\ast}) ^{2} - m _{\ast} ^{2}}$.  

{ Having obtained a closed expression for $T(E,k_y)$, we now analyze its dependence on barrier height, incidence angle, and slab thickness. The representative material parameter ranges employed below are chosen to be consistent with the experimentally motivated heterostructure scales discussed in Sec.~\ref{QWZ_model} for magnetic topological-insulator systems based on (Bi,Sb)$_2$Te$_3$.} {Electrostatic potentials of the order of several tens to a few hundreds of meV can be applied through local gate voltages, providing control over the energy landscape experienced by Dirac carriers \cite{cho_gate-tunable_2011}. In particular, gate-induced potential steps $V_{\ast} \sim 100$ - $200$~meV allow one to access distinct transport regimes \cite{katsnelson_chiral_2006}. In addition, the width of the topological region can be engineered within the range $L \sim 20$ - $100$~nm, which sets the relevant interference and tunneling length scales. These experimentally realistic values motivate our choice of dimensionless parameters and justify the scaling adopted throughout the following analysis \cite{hasan_colloquium_2010, qi_topological_2011}. }

\begin{figure}[t]
    \centering
    \includegraphics[trim={2.6cm 1.5cm 1.4cm 0.9cm},width=0.6\textwidth]{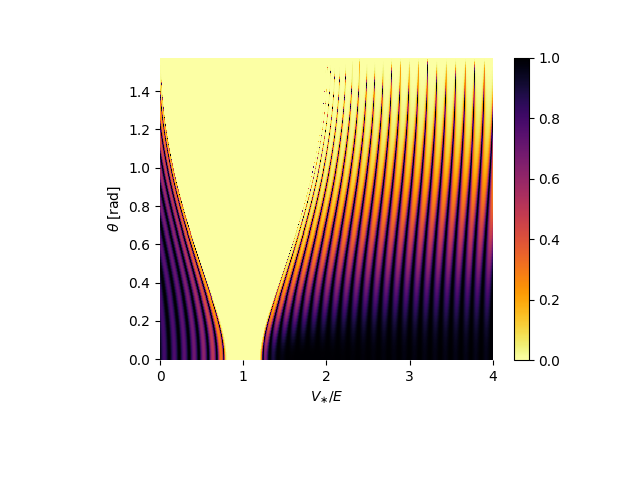}
    \caption{Angle- and barrier-dependent transmission probability from Eq.~(\ref{Transmission1}). The color scale represents the transmission probability $T$, ranging from 0 (yellow, fully suppressed transmission) to 1 (dark purple, perfect transmission). For $V_{*}/E \lesssim 1$ propagating modes produce Fabry-Pérot fringes with resonances at $q_{x}L=n\pi$. For $V_{*}/E \gtrsim 1$ the modes become evanescent, leading to exponential suppression with narrow tunneling windows near the propagating-evanescent boundary.}
    \label{TransmissionMap}
\end{figure}

Figure~\ref{TransmissionMap} summarizes the main features of the transmission probability across the trivial-topological-trivial junction. For $V_{*}/E \lesssim 1$, the longitudinal momentum inside the slab is real and propagating states give rise to Fabry-Pérot oscillations, consistent with earlier analyses of quantum interference in Dirac materials~\cite{katsnelson_chiral_2006,Beenakker2008}.  In this regime, transmission is maximized at normal incidence due to pseudospin locking and suppressed at large incidence angles, reflecting the momentum-selective nature of transport in Chern phases. { The dark ridges visible in the density plot correspond to perfect-transmission resonances at $q_{x}L=n\pi$, while bright troughs appear at antiresonances when $q_{x}L=(n+1/2)\pi$.}

For $V_{*}/E \gtrsim 1$, the longitudinal momentum becomes imaginary and the states inside the slab turn evanescent. Transmission then decays exponentially with barrier thickness, in agreement with the standard tunneling form~\cite{datta_electronic_1995}. Only narrow angular windows remain open close to the propagating-evanescent boundary, where $\kappa L$ is small, leading to sharp resonant features. This crossover from propagating-mode interference to tunneling is further controlled by the interface matching factor $\Gamma_{ss_{\ast}}$, which modulates the overall amplitude. Altogether, the results highlight the robustness of near-normal-incidence transmission in topological systems~\cite{Qi2006,Qiao2014} 
and demonstrate that perfect and near-perfect transmission arises from the spinor matching enabled by band inversion across the junction, encoded in the sign reversal of the Dirac mass in the central region. Within the Qi-Wu-Zhang Chern-insulator model, this mass inversion corresponds to the transition between trivial and topological phases. Beyond the transmission itself, Berry curvature plays a decisive role in the transverse and nonlinear transport responses analyzed below, where it enters explicitly in the Hall conductance through the anomalous velocity term.

\section{Nonlinear conductance expansion within the Landauer framework} \label{longitudinal_conductance_section}

In this work, we employ the Landauer formalism to investigate electronic transport through heterostructures based on Chern insulators \cite{landauer_electrical_1957, buttiker_four-terminal_1986, datta_electronic_1995}. The steady-state current flowing through such phase-coherent systems connected to two electronic reservoirs is expressed as an energy integral over the difference of the Fermi-Dirac distributions of the contacts, weighted by the energy-dependent transmission probability of the heterostructure. To model experimentally relevant conditions, we adopt an asymmetric bias configuration in which the drain reservoir is held at the equilibrium chemical potential $\mu _{D} = E _{F}$, while a bias voltage $V$ is applied entirely to the source, shifting its chemical potential to $\mu _{S} = E _{F} + eV$. Under this setup, the Landauer expression for the current reads
\begin{align}
    I = \frac{2e}{h} \int _{-\infty} ^{\infty} T (E) \, \left[f(E-eV)-f(E) \right] \, d E , 
    \label{Landauer_formula}
\end{align}
where $T(E)$ is the transmission probability for electrons of energy $E$, and $f(E) = \left[ 1 + e ^{ (E-E _{F}) / k _{B} T} \right] ^{-1}$ is the Fermi-Dirac distribution at temperature $T$. The prefactor $2e/h$ accounts for spin degeneracy; each spin channel contributes independently and equally to the conductance in the absence of spin-selective mechanisms.

To analyze the response of the system at small bias, we expand the current in powers of $V$. Since the bias enters only through the source distribution, $f(E-eV)$, we perform a Taylor expansion around $V=0$ \cite{kawabata_nonlinear_2022}:
\begin{align}
    f(E-eV) = f (E) - eV f'(E) + \frac{(eV) ^{2}}{2} f''(E) - \frac{(eV) ^{3}}{6} f'''(E) + \mathcal{O} (V ^{4}) .  \label{Expansion_FD_distribution} 
\end{align}
Substituting into the Landauer formula yields
\begin{align}
    I (V) = G _{1} V + G _{2} V ^{2} + G _{3} V ^{3} + \mathcal{O} (V ^{4}) 
\end{align}
where the transport coefficients are given by
\begin{align}
    G _{n} (\mu ) =  \frac{2e ^{n+1}}{h} \frac{(-1) ^{n}}{n!} \int _{-\infty} ^{\infty} T(E) \, \frac{d ^{n} f}{dE ^{n}} \, d E , \quad n \geq 1 . 
\end{align}
We now integrate by parts the expression for $G_{n}$, in order to transfer the energy derivatives from the Fermi-Dirac distribution to the transmission function. This transformation is justified under the physical assumption that the transmission $T(E)$ is a smooth function of energy that remains bounded as $ \vert E \vert \to \infty$, while the Fermi function and all its derivatives decay exponentially in the same limit. As a result, the boundary terms generated by the integration by parts vanish identically. Thus we obtain
\begin{align}
    G _{n} (\mu ) =  \frac{2e ^{n+1}}{h} \frac{1}{n!} \int _{-\infty} ^{\infty}   \frac{d ^{n-1} T(E)}{dE ^{n-1}} \, \left(- \frac{d f (E)}{dE } \right) \, d E , \quad n \geq 1 . \label{General_longitudinal_conductance}
\end{align}
In this form, the transport coefficients are seen to be controlled by successive energy derivatives of the transmission function, evaluated within the thermal window around the Fermi energy where the Fermi function is appreciably different from zero. Here, $G _{1}$ corresponds to the linear conductance and is sensitive to the value of the transmission function near the Fermi energy, within an energy window of width $\sim k _{B}T$. In the limit where $T(E)$ varies slowly around $E_{F}$, this reduces to $G_{1} \approx (2e ^{2} / h) T(E_{F})$. The quadratic coefficient $G _{2}$, which is even in $V$, reflects the leading-order nonlinear correction and encodes nonreciprocal transport: a deviation from the antisymmetric behavior $I(-V) = -I(V)$. Finally, the cubic term $G _{3}$ captures the first odd-order nonlinear contribution and becomes particularly relevant in the vicinity of sharp features in the transmission spectrum, such as resonant states or band edges. The choice of an asymmetric voltage drop, where the full bias is applied to the source while the drain remains grounded, is essential for capturing the full structure of the current-voltage response, including both reciprocal and nonreciprocal components.

Having established the general framework for conventional electronic transport using the Landauer formalism, we now apply this theory to analyze the current response of the heterostructure introduced in the previous section, which consists of a central Chern insulator strip connected on both sides to leads made of topologically trivial material. This setup defines a junction between trivial and nontrivial phases. 
By evaluating the energy-dependent transmission function $T(E)$ of this heterostructure, we aim to characterize the linear and nonlinear components of the conventional current and to explore how topological features of the central region manifest in the bias-dependent transport response. To this end, some considerations are in order. Let $L_{x}$ and $L _{y}$ denote the lengths of the system along the $x$- and $y$-directions, respectively. The system extends along $x$ in the region $\vert x \vert \leq L _{x}/2 $, and is connected to two leads at $\vert x \vert \geq L _{x}/2$. These leads are assumed to be made of graphene and subjected to a large potential $V_{0}$ and the central potential is assumed to be at zero potential $V_{\ast} = 0$. In the limit $\vert V_{0} \vert \to \infty$, an infinite number of propagating modes exist in the leads. Along the $y$-direction, appropriate boundary conditions are imposed, allowing the use of a Fourier representation with transverse wave number $k _{y}$. Since we are ultimately interested in the limit $L _{y} \to \infty$, the specific choice of boundary conditions along $y$ becomes irrelevant for the transport properties of the system.

Accordingly, the expression for the transmission probability in Eq. (\ref{Transmission1}) reduces to 
\begin{align}
    T (E,k_{y}) = \frac{ 1 }{  \cos ^{2} (q _{x} L _{x}) +  \left( \frac{E}{ \hbar v _{F}  q _{x} }  \right) ^{2} \, \sin ^{2} (q _{x} L _{x})  } , \label{Transmission}
\end{align}
where $ q _{x} = \frac{1}{\hbar v _{F} } \sqrt{  E ^{2} - m _{\ast} ^{2} - ( \hbar v _{F} k _{y} ) ^{2} }$ is the wave number along the $x$ direction, which depends on the energy $E$ and the wave number $k _{y}$ along the $y$ direction. In order to obtain the energy-dependent transmission function $T(E)$, we have to consider all the modes along the $y$ direction. In the limit $L_{y} \to \infty$, we have \cite{datta_electronic_1995, buttiker_quantized_1990}
\begin{align}
    T(E) = \sum _{k _{y}} T (E,k_{y}) \, \to \, \frac{L_{y}}{\pi} \int _{0} ^{\infty} T (E,k_{y}) \, dk_{y} . 
\end{align}
To simplify the expression further, we perform a change of variables by introducing $\kappa _{y} = k _{y} L _{x}$, and rewrite the result in terms of the dimensionless quantities $\epsilon = \frac{ E L _{x} }{\hbar v _{F} } $ and $\lambda = \frac{ m _{\ast} L _{x} }{\hbar v _{F} }$. We obtain
\begin{align}
    T( \epsilon ) = \frac{L_{y}}{\pi L _{x} } \int _{0} ^{\infty} \frac{ d \kappa _{y} }{  \cos ^{2} \xi + (\epsilon / \xi ) ^{2} \; \sin ^{2} \xi  } . 
\end{align}
where $ \xi (\kappa _{y}) =  \sqrt{ \epsilon ^{2} - \lambda ^{2} - \kappa _{y} ^{2} }$. A key distinction arises in the transmission probability, which changes qualitatively depending on the nature of $\xi (\kappa _{y})$. For $\kappa _{y}$ below the threshold $\sqrt{\epsilon ^{2} - \lambda ^{2}}$, $\xi (\kappa _{y})$ remains real, while for larger $\kappa _{y}$, it becomes imaginary. This results in markedly different behaviors in each regime:
\begin{align}
    T( \epsilon ) = \frac{L_{y}}{\pi L _{x} } \left[  \int _{0} ^{\sqrt{\epsilon ^{2} - \lambda ^{2}}} \frac{  d \kappa _{y} }{  \cos ^{2} \xi +   (\epsilon / \xi ) ^{2} \; \sin ^{2} \xi  }   +  \int _{\sqrt{\epsilon ^{2} - \lambda ^{2}}} ^{\infty} \frac{  d \kappa _{y} }{  \cosh ^{2} \xi +  (\epsilon / \xi ) ^{2} \; \sinh ^{2} \xi  }  \right] . 
\end{align}
In this expression, the first contribution corresponds to transport through propagating modes, where the longitudinal momentum remains real and describes standard scattering processes. In contrast, the second contribution originates from evanescent states with imaginary longitudinal momentum, which capture tunneling phenomena across the heterostructure.

To simplify the transmission integral it is convenient to eliminate the square-root factors in both contributions by suitable changes of variables. In the propagating sector, where the integration runs from zero to $\sqrt{\epsilon ^{2} - \lambda ^{2}}$, we introduce the substitution $\kappa _{y} = \Delta \sin \theta $, with $\Delta = \sqrt{\epsilon ^{2} - \lambda ^{2}}$ and $\theta \in [0, \pi /2]$. This change maps the semi-infinite range to a compact interval and transforms the measure according to $d \kappa _{y} = \Delta \cos \theta d \theta $ and $\sqrt{\epsilon ^{2} - \lambda ^{2} - \kappa _{y} ^{2} } = \Delta \cos \theta $. The result is a smooth integrand expressed entirely in terms of $\theta$. In the evanescent sector, corresponding to the second integral over $[0, \infty )$, we use instead the substitution $\kappa _{y} = \Delta \sinh u$, with $u \in [0, \infty ) $. In this case one has $d \kappa _{y} = \Delta \cosh u \, du$ and $\sqrt{\epsilon ^{2} - \lambda ^{2} + \kappa _{y} ^{2} } = \Delta \cosh u$. This transformation removes the square root and yields an integrand that decays rapidly at large $u$. With both substitutions, the transmission can be written in the compact form 
\begin{align}
    T( \epsilon ) = \frac{L_{y}}{\pi L _{x} } \left[  \int _{0} ^{ \pi /2 } \frac{ \Delta  \sin \theta \, d \theta  }{  \cos ^{2} (\Delta   \sin \theta ) +   (\epsilon / \Delta ) ^{2}  \csc ^{2} \theta   \sin ^{2} (\Delta \, \sin \theta) } + \int _{0} ^{\infty} \frac{ \Delta  \sinh u \, du }{  \cosh ^{2} ( \Delta  \sinh u ) +  (\epsilon / \Delta ) ^{2}   \mbox{csch } ^{2} u \, \sinh ^{2} ( \Delta  \sinh u )  }   \right] . \label{transmission_probability_integrals} 
\end{align}
The transformed integrals presented above no longer admit a closed-form evaluation in terms of elementary functions. Nevertheless, they are now in a form that is well suited for direct numerical computation on finite and stable domains. Beyond the numerical evaluation, it is also instructive to analyze the behavior of the transmission in asymptotic regimes of the parameter $ \Delta = \sqrt{\epsilon ^{2} - \lambda ^{2}}$. 

For $\Delta \gg 1$, the trigonometric part of the integrand oscillates rapidly, and one may replace $\sin ^{2} (\Delta \, \sin \theta)$ and $\cos ^{2} (\Delta \, \sin \theta)$ by their average values of $1/2$. Under this approximation, the transmission approaches the asymptotic form
\begin{align}
    T( \epsilon ) \simeq \frac{L_{y}}{ L _{x} }   \left( \frac{\epsilon}{\sqrt{2\epsilon ^{2} - \lambda ^{2}}} - 1 \right)  + \mathcal{O} (\Delta ^{-1/2}) , 
\end{align}
which captures the quasi-classical limit where the rapid oscillations are effectively averaged out.

In contrast, for $\Delta \ll 1$, the arguments of the trigonometric and hyperbolic functions are small, so that 
$\sin (\Delta \, \sin \theta) \approx \Delta \, \sin \theta$, $\cos (\Delta \, \sin \theta) \approx 1 $, and similarly for the hyperbolic sector. Expanding to leading order, the transmission reduces to
\begin{align}
    T( \epsilon ) \simeq \frac{L_{y}}{ 2L _{x} }  \frac{\epsilon ^{2} + \lambda ^{2}}{(1 + \epsilon ^{2}) ^{2}} + \mathcal{O} (\Delta ^{2}) , 
\end{align}
which remains finite as $\Delta \to 0 $. This expression highlights the perturbative regime in which the response is dominated by the smooth low-energy expansion. Taken together, these limiting forms provide analytic benchmarks that complement the full numerical evaluation of the transmission, and they help elucidate the crossover between the oscillatory, large-$\Delta$ regime and the regular, small-$\Delta$ regime.

\begin{figure}
    \centering
    \includegraphics[width=0.44\linewidth]{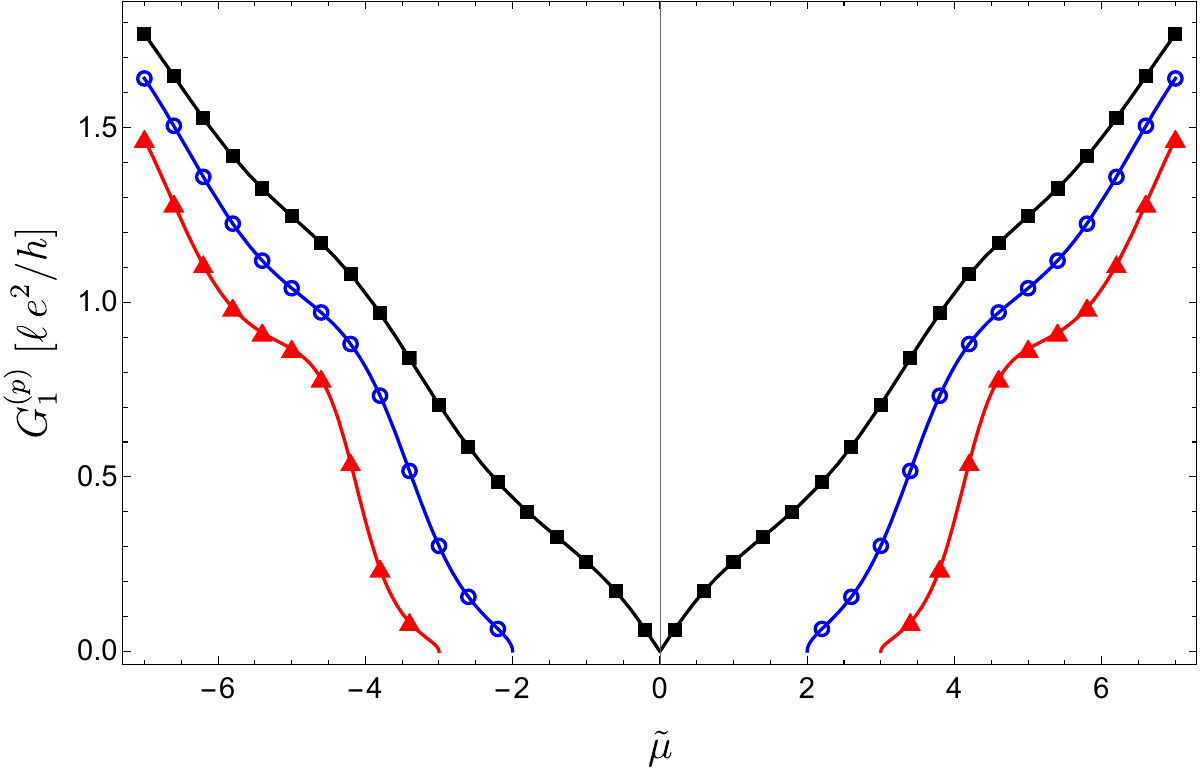} \hspace{0.5cm}
    \includegraphics[width=0.44\linewidth]{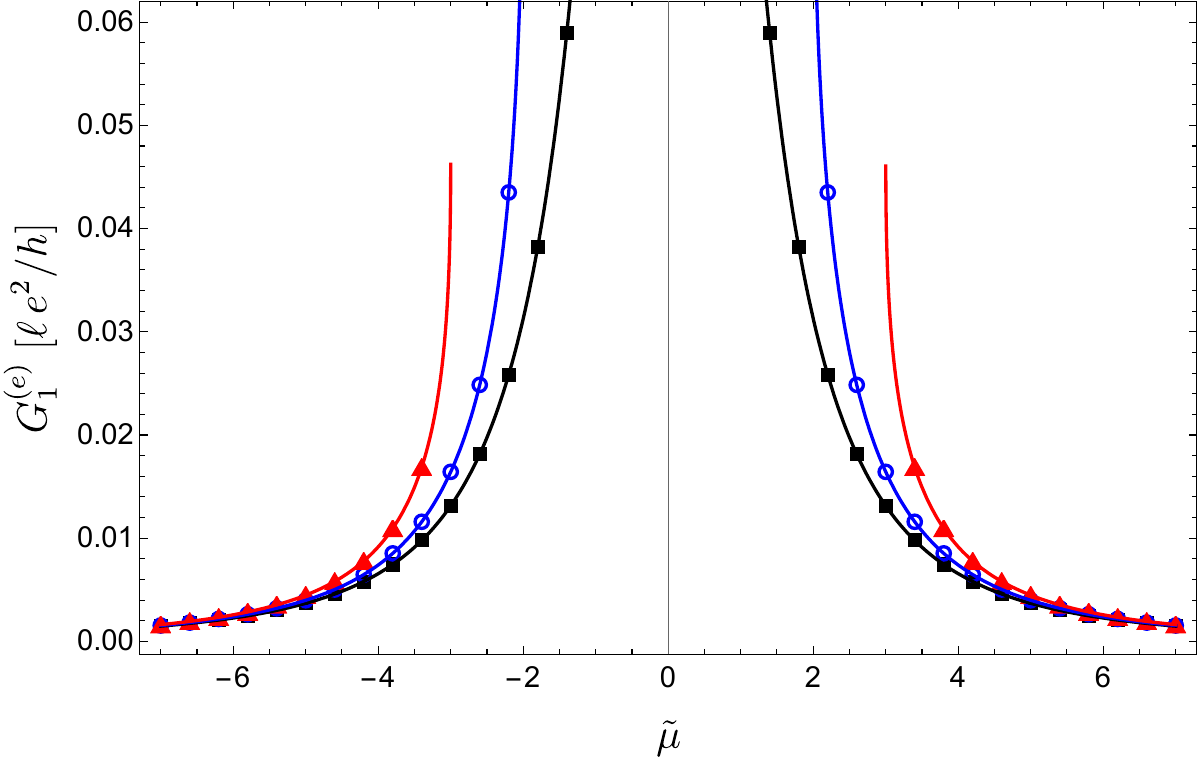}\\
    \includegraphics[width=0.44\linewidth]{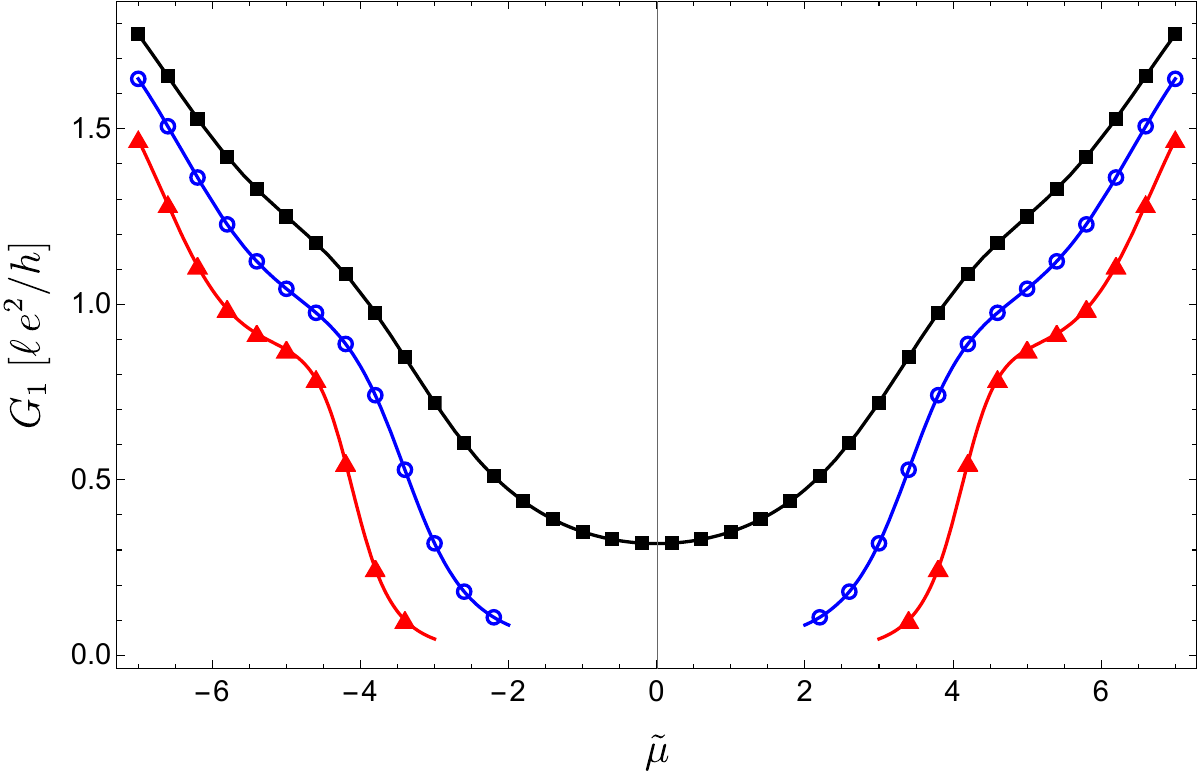}
    \caption{Linear longitudinal conductance, in units of $\ell e^{2}/h$ with $\ell=L_{y}/L_{x}$, as a function of the rescaled chemical potential $\tilde{\mu} \equiv \frac{ \mu L _{x} }{\hbar v _{F} }$. Top left: contribution from propagating modes $G_{1}^{(p)}$, associated with real longitudinal momentum. Top right: contribution from evanescent modes $G_{1}^{(e)}$, arising from imaginary longitudinal momentum and tunneling transport. Bottom: total conductance $G_{1} = G_{1}^{(p)} + G_{1}^{(e)}$ for different gap values, with $\lambda=0$ (black squares), $\lambda=1$ (blue circles), and $\lambda=2$ (red triangles).}
    \label{longitudinal_linear_conductance}
\end{figure}

The conductance coefficients can be systematically obtained from the expansion of the Landauer formula in powers of the applied bias. The first-order term corresponds to the linear conductance, which is directly proportional to the total transmission probability of the junction,
\begin{align}
    G _{1} ( \tilde{\mu} ) = \frac{e ^{2} }{h} \, T ( \tilde{\mu} ) , \label{linear_longitudinal_conductance}
\end{align}
where $\tilde{\mu} \equiv \frac{ \mu L _{x} }{\hbar v _{F} }$ is a rescaled chemical potential. This expression highlights the fact that the linear response is entirely governed by the probability of electron transmission across the heterostructure. In order to illustrate the different transport channels contributing to the linear longitudinal conductance (\ref{linear_longitudinal_conductance}), we present three panels in Fig. \ref{longitudinal_linear_conductance}. The top left panel shows the conductance (in units of $\ell e ^{2}/h$, with the geometric factor $\ell = L _{y}/L _{x}$) arising solely from propagating modes, where the longitudinal momentum is real and transport is mediated by extended states. This contribution, denoted as $G _{1} ^{(p)} ( \tilde{\mu} )$, corresponds to the first integral term in Eq. (\ref{transmission_probability_integrals}). The top right panel isolates the contribution of evanescent modes (again in units of $\ell e ^{2}/h$), associated with imaginary longitudinal momentum and therefore describing tunneling processes. This part, denoted as $G _{1} ^{(e)} ( \tilde{\mu} )$, follows directly from the second integral in Eq. (\ref{transmission_probability_integrals}). Finally, the bottom panel presents the total conductance $G _{1} ( \tilde{\mu} )  = G _{1} ^{(p)} ( \tilde{\mu} ) + G _{1} ^{(e)} ( \tilde{\mu} )$ as a function of the rescaled chemical potential $\tilde{\mu}$. The black curve with square markers corresponds to the gapless case ($\lambda=0$), representative of graphene, where the conductance decreases smoothly toward a minimum around charge neutrality and grows symmetrically for electron and hole doping. When a finite mass term is introduced, as in the Chern insulating phases, the conductance is suppressed within the gap region and the overall magnitude is reduced. This effect is illustrated by the blue curve with circles for $\lambda=1$ and by the red curve with triangles for $\lambda=2$, where the larger gap leads to a more pronounced flattening near $\tilde{\mu}=0$ and a delayed onset of conduction as $\tilde{\mu}$ increases. The comparison highlights how the opening of a topological gap systematically diminishes the available propagating channels and shifts transport toward regimes dominated by evanescent contributions.

Beyond the linear regime, nonlinear corrections appear in the form of higher-order derivatives of the transmission with respect to energy, evaluated at the chemical potential. In particular, the second- and third-order contributions take the form
\begin{align}
    G _{2} ( \tilde{\mu}) = \frac{e ^{3} }{2h} \,  \frac{ L _{x} }{\hbar v _{F} } \frac{d \, T ( \epsilon )}{d \epsilon } \Bigg| _{\epsilon = \tilde{\mu} } , \qquad G _{3} ( \tilde{\mu}) = \frac{e ^{4} }{6h} \,  \frac{ L _{x} ^{2} }{ ( \hbar v _{F} ) ^{2} } \frac{d ^{2} \, T ( \epsilon )}{d \epsilon ^{2} } \Bigg| _{\epsilon = \tilde{\mu} }  .
\end{align}
These expressions make explicit that nonlinear transport is governed not only by the absolute value of the transmission, but also by its energy dependence around the Fermi level. The derivatives encode the sensitivity of the transmission to changes in the incoming energy, thus determining the strength of quadratic and cubic conductance contributions.

\begin{figure}
    \centering
    \includegraphics[width=0.47\linewidth]{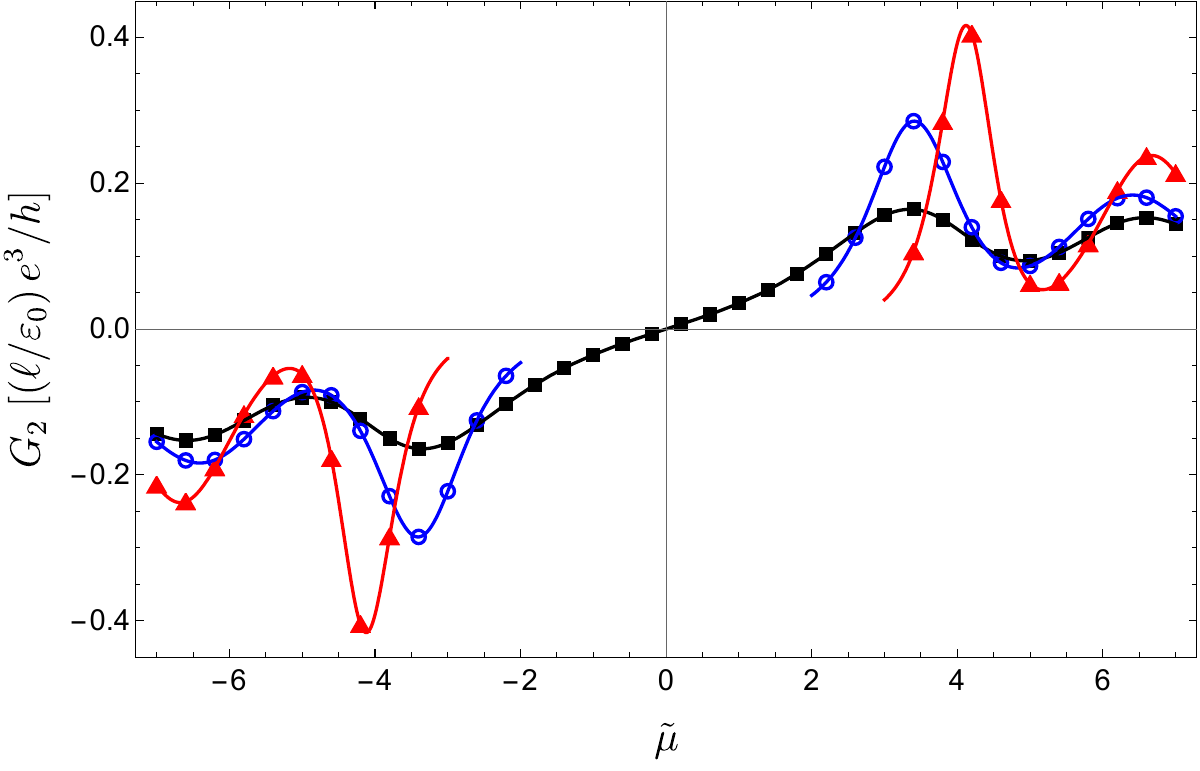} \;\;\;\; \includegraphics[width=0.47\linewidth]{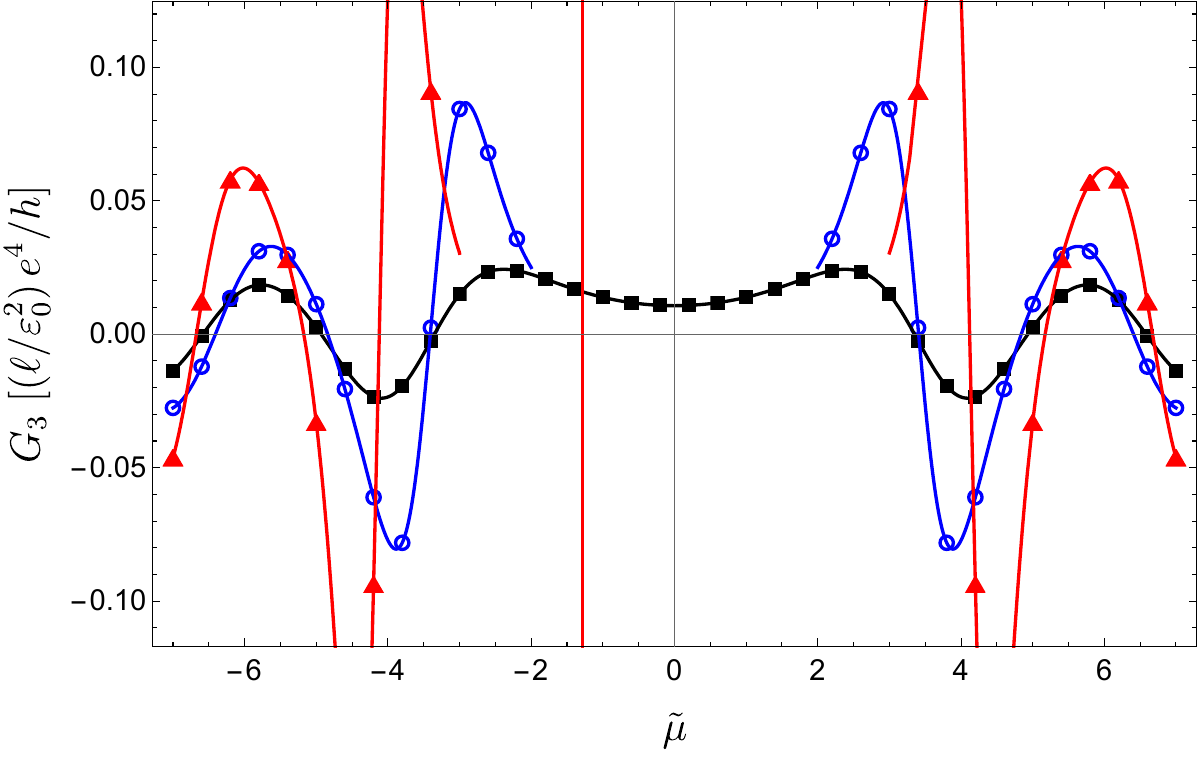}
    \caption{Nonlinear longitudinal conductances as functions of the rescaled chemical potential $\tilde{\mu}$. Left: second-order conductance $G_{2}$ (in units of $(\ell/\varepsilon_{0})e^{3}/h$), showing a smooth evolution for the gapless case ($\lambda=0$, black squares) and pronounced oscillations with sign reversals for finite gaps ($\lambda=1$, blue circles; $\lambda=2$, red triangles). Right: third-order conductance $G_{3}$ (in units of $(\ell/\varepsilon_{0}^{2})e^{4}/h$), which remains weak and nearly featureless in the gapless limit but develops sharp peaks and strong oscillations in the Chern insulating phases.}
    \label{nonlinear_conductances}
\end{figure}

The left and right panels of Fig. \ref{nonlinear_conductances} show the nonlinear longitudinal conductances $G_{2}$ (in units of $ (\ell / \varepsilon _{0}) e ^{2}/h$) and $G_{3}$ (in units of $ (\ell / \varepsilon _{0} ^{2}) e ^{2}/h$), respectively, as functions of the rescaled chemical potential $\tilde{\mu} = \mu / \varepsilon _{0}$, where $\varepsilon _{0} = \hbar v _{F} / L_{x}$ is a characteristic energy scale. The second-order conductance $G_{2}$ displays a relatively smooth and monotonic evolution in the gapless case ($\lambda=0$), crossing from negative values under hole doping to positive values under electron doping. When a finite mass is introduced ($\lambda=1,2$), the response acquires pronounced oscillations and sign reversals, particularly near the band edges, signaling the growing influence of tunneling and interference effects in the nonlinear regime. In contrast, the third-order conductance $G_{3}$ exhibits a qualitatively different behavior: while the graphene limit shows only weak fluctuations around zero, the Chern insulating phases develop strong, highly structured oscillations with sharp peaks and sign changes. These features highlight the enhanced sensitivity of higher-order transport to the details of the band structure and to the presence of a topological gap. Taken together, the results indicate that the opening of a mass gap not only suppresses the linear conductance but also amplifies nonlinear responses, with higher-order conductivities providing a clear fingerprint of topological band modifications. We note that the influence of dephasing and disorder on these Fabry-Pérot oscillations will be addressed separately in Sec.~VI, where it is shown that the main features remain robust once phase coherence is partially lost.

\section{Expansion of the Hall conductance: Linear and nonlinear contributions} \label{Hall_conductance_section}

Having established the longitudinal response, we now turn to the transverse conductance, i.e., the Hall channel, within the Landauer framework. At the macroscopic scale, the quantum Hall conductance is well known to be quantized in units of $e ^{2}/h$, with the quantization directly proportional to the Chern number of the occupied bands \cite{thouless_quantized_1982, hatsugai_chern_1993}. In mesoscopic heterostructures, however, the situation is richer: nonlinear contributions to the Hall current can arise due to the interplay between the applied bias and the potential landscape, leading to deviations from perfect quantization \cite{sodemann_quantum_2015, PhysRevB.98.155125}. In the fully quantum regime, where interference effects are prominent, the transverse response is influenced by the position of the chemical potential relative to the band edges and the available transmission channels. Within the Landauer expansion, these effects manifest as higher-order terms in the bias, providing access to the nonlinear Hall conductance \cite{buttiker_four-terminal_1986, datta_electronic_1995}. In what follows, we develop this expansion and analyze how Berry curvature, mass inversion, and electrostatic barriers conspire to generate both the linear and nonlinear transverse conductance. While the transmission properties discussed in Sec.~\ref{Klein_tunneling_section} are governed by the band inversion encoded in the mass profile, the transverse response analyzed in this section depends explicitly on the Berry curvature of the Chern-insulating region. In contrast to the longitudinal transmission, which follows from the matching of Dirac spinors across the junction, the Hall response provides a direct transport manifestation of the topological band structure, as it originates from the anomalous velocity associated with Berry curvature.

Next we closely follow Ref. \cite{kawabata_nonlinear_2022}. Let us consider a two-dimensional sample of size $L_{x} \times L_{y}$, connected to two electronic reservoirs. A bias voltage $V$, is applied along the $x$ direction, such that the left reservoir is described by the Fermi distribution $f(E-eV)$, while the right reservoir remains at equilibrium with distribution $f(E)$. Along the transverse $y$-direction, we assume periodic boundary conditions, so that no edge effects interfere with the analysis. This setup allows us to treat the transport problem as a scattering process between left and right contacts, with transmission probability $T(\mathbf{k})$ depending on momentum.

The current carried by an electron with momentum $\mathbf{k}$ has two contributions. The first is the conventional group velocity term $\nabla _{\mathbf{k}} E (\mathbf{k}) / \hbar$, which accounts for transport along the applied bias. The second is the anomalous velocity term,
\begin{align}
    \mathbf{v} _{\mbox{\scriptsize anom}} = - \frac{e}{\hbar} \, \mathbf{E} \times \boldsymbol{\Omega} (\mathbf{k}) ,
\end{align}
which originates from the Berry curvature $\boldsymbol{\Omega} (\mathbf{k})$ \cite{xiao_berry_2010}. This anomalous velocity points along the transverse direction and is therefore responsible for the Hall current \cite{jungwirth_anomalous_2002}. In the absence of Berry curvature (trivial bands), this contribution vanishes.

A crucial ingredient is the definition of the electric field inside the conductor. In the semiclassical picture, one would take $E_{x}=V/L_{x}$. However, in the coherent transport regime described by Landauer theory, the voltage drop across the system depends on the transmission. If $T=1$, the system is perfectly transmitting, and no potential drop develops inside the sample: the entire bias is absorbed at the contacts. Conversely, for $T \ll 1$, the sample is almost opaque, and the field approaches the semiclassical value $V/L_{x}$. In general, the effective electric field inside the system is given by \cite{datta_electronic_1995, Imry2002}
\begin{align}
    E_{x} = (1-T) \frac{V}{L_{x}} ,
\end{align}
which naturally interpolates between the coherent and semiclassical limits. This factor is the origin of the additional $1-T$ that multiplies the transmission probability in the final expression for the Hall current.

To compute the total current, we evaluate the imbalance between modes transmitted from left to right and from right to left, each weighted by the transmission probability and the corresponding Fermi occupation. The number of transmitted states in the momentum range $d ^{2} \mathbf{k}$ is proportional to
\begin{align}
    dN_{L \to R} \sim T (\mathbf{k}) \, f(E-eV) \, \frac{L_{y} \, d ^{2} \mathbf{k} }{(2 \pi ) ^{2}} , \qquad dN_{R \to L} \sim T (\mathbf{k}) \, f(E) \, \frac{L_{y} \, d ^{2} \mathbf{k} }{(2 \pi ) ^{2}} . 
\end{align}
Subtracting these contributions and multiplying by the anomalous velocity yields the differential Hall current
\begin{align}
    dI _{H} = \frac{e ^{2}}{\hbar} \frac{L _{y}}{L _{x}} V \,  T (\mathbf{k}) \, [1- T (\mathbf{k})] \, \Omega (\mathbf{k}) \, [f(E-eV) - f(E) ] \, \frac{ d ^{2} \mathbf{k} }{(2 \pi ) ^{2}} ,
\end{align}
where $\Omega (\mathbf{k})$ is the only nonzero component of the Berry curvature in two-dimensional systems. The dependence on the bias enters through the difference of Fermi functions. Expanding this quantity in powers of $V$, one finds that the linear contribution cancels out due to symmetry. Therefore, the leading Hall term arises at quadratic order in the applied bias. This motivates the definition of the nonlinear Hall conductance, obtained by collecting the coefficient of $V ^{2}$. The result is
\begin{align}
    G_{2} ^{\mbox{\scriptsize H}} (\mu ) = - \frac{e ^{3}}{\hbar} \, \frac{L_{y}}{L_{x}} \, \int _{\mbox{\scriptsize BZ}} \frac{ d ^{2} \mathbf{k} }{(2 \pi ) ^{2}} \, T (\mathbf{k}) \, [1- T (\mathbf{k})] \, \Omega (\mathbf{k}) \, \left( - \frac{\partial f(E) }{\partial E} \right) , \label{2nd_order_Hall_conductance}
\end{align}
where the momentum integral is taken over the whole Brillouin zone. At finite temperature, the derivative of the Fermi function smoothly selects states near the chemical potential. At zero temperature, it reduces to a delta function, so that only states at the Fermi surface $E (\mathbf{k}) = \mu $ contribute. This reveals the Fermi-liquid character of the  nonlinear Hall response: it depends solely on the Berry  curvature evaluated on the Fermi contour. Finally, the prefactor $T(1-T)$ encodes the quantum-coherent origin of the effect, since the response vanishes in both the tunneling  limit $T \ll 1$ and in the ballistic limit $T=1$.

Using the series expansion of the Fermi-Dirac distribution, given by Eq. (\ref{Expansion_FD_distribution}), it is straightforward to generalize the above derivation to higher orders in the applied bias. Each term in this expansion yields an additional nonlinear contribution to the Hall conductance, so that the total transverse response can be expressed as a hierarchy of coefficients $G_{n} ^{\mbox{\scriptsize H}} (\mu ) $ multiplying powers of the voltage. While the leading term vanishes and the quadratic contribution $G_{2} ^{\mbox{\scriptsize H}} (\mu ) $ dominates at low bias, the higher-order terms encode further corrections arising from quantum coherence and the detailed energy dependence of the transmission probability and the Berry curvature. A general formula for the higher-order terms is $G_{n} ^{\mbox{\scriptsize H}} (\mu ) = \frac{e ^{n-2}}{(n-1)!} \frac{\partial ^{n-2}}{\partial \mu ^{n-2}} G_{2} ^{\mbox{\scriptsize H}} (\mu )$. This systematic expansion thus provides a unified framework for analyzing both the second-order and higher-order nonlinear Hall effects within the scattering formalism.

Building on the general scattering framework for the nonlinear Hall response, we now specialize our analysis to the heterostructure introduced in the previous section. The system under consideration consists of a central Chern insulating strip, subjected to an electrostatic potential barrier, and flanked by topologically trivial leads. This configuration realizes a junction between trivial and nontrivial phases, where the mass inversion and the presence of Berry curvature provide the microscopic origin of the transverse current. In the following, we evaluate the Hall conductance of this setup by applying the Landauer expansion derived above, focusing on how the transmission properties of bulk states and the electrostatic profile conspire to generate nonlinear transverse transport.

In the problem at hand, the nonzero component of the Berry curvature is $\boldsymbol{\Omega} (\mathbf{k}) = \Omega (k) \,  \hat{\mathbf{z}}$ \cite{qi_topological_2006}, where
\begin{align}
    \Omega (k) = - s \frac{m _{\ast} \, (\hbar v _{F}) ^{2}}{2 \, [(\hbar v _{F} k) ^{2} + m _{\ast} ^{2} ] ^{3/2}} . \label{Berry_curvature} 
\end{align}
We now proceed to evaluate the scattering expression for the second-order Hall conductance [Eq. (\ref{2nd_order_Hall_conductance})] by inserting the explicit form of the Berry curvature [Eq. (\ref{Berry_curvature})] together with the transmission probability $T (\mathbf{k})$ [Eq. (\ref{Transmission})]. After performing the change of variables $\kappa _{x} = k _{x} L _{x}$ and $\kappa _{y} = k _{y} L _{x}$, and introducing the dimensionless variables $\epsilon = \frac{ E L _{x} }{\hbar v _{F} } $ and $\lambda = \frac{ m _{\ast} L _{x} }{\hbar v _{F} }$, the expression reduces to
\begin{align}
    G_{2} ^{\mbox{\scriptsize H}} (\mu ) =   \frac{e ^{3}}{2 \hbar} \, \frac{L_{y}}{\hbar v _{F}}  \, \int _{- \infty} ^{\infty} \int _{- \infty} ^{\infty} \frac{ d \kappa _{x} \, d \kappa _{y} }{(2 \pi ) ^{2}} \,  \frac{  [  \left( \epsilon / \kappa _{x}   \right) ^{2} - 1 ] \, \sin ^{2} \kappa _{x}  }{  [ \cos ^{2} \kappa _{x} +  \left( \epsilon / \kappa _{x}   \right) ^{2} \, \sin ^{2} \kappa _{x}  ] ^{2} } \,  \frac{ s \lambda }{  (  \kappa ^{2} +  \lambda ^{2} ) ^{3/2}}  \, \delta \left( \frac{\mu L _{x}}{\hbar v _{F}} - s   \sqrt{ \kappa ^{2}  + \lambda ^{2} } \right) , \label{Integral_Hall}
\end{align}
where $ \kappa _{x} (\kappa _{y}) \equiv  \sqrt{ \epsilon ^{2} - \lambda ^{2} - \kappa _{y} ^{2} }$. It is worth noting that the location of the chemical potential $\mu$ determines which band contributes to transport, thereby fixing the band index $s$. For convenience, we encode this by writing $s = \mbox{sgn}(\mu)$, so that the chemical potential can be expressed as $\mu = s \vert \mu \vert $. Introducing the dimensionless chemical potential $\tilde{\mu} = \frac{ \vert \mu \vert L _{x}}{\hbar v _{F}}$, the integral (\ref{Integral_Hall}) simplifies to
\begin{align}
    G_{2} ^{\mbox{\scriptsize H}} (\mu ) =   \frac{  e ^{3}}{ \pi h  } \, \frac{ \lambda L_{y}}{\hbar v _{F}} \, \frac{\mbox{sgn}(\mu)}{\tilde{\mu} ^{3} } \,   \int _{0} ^{\infty} \int _{0} ^{\infty}  d \kappa _{x} \, d \kappa _{y} \,  \frac{  [  \left( \tilde{\mu} / \kappa _{x}   \right) ^{2} - 1 ] \, \sin ^{2} \kappa _{x} }{  [ \cos ^{2} \kappa _{x} +  \left( \tilde{\mu} / \kappa _{x}   \right) ^{2} \, \sin ^{2} \kappa _{x}  ] ^{2} } \; \delta \left( \tilde{\mu} - \sqrt{ \kappa ^{2}  + \lambda ^{2} } \right) ,
\end{align}
where $ \kappa _{x} (\kappa _{y}) \equiv  \sqrt{ \tilde{ \mu } ^{2} - \lambda ^{2} - \kappa _{y} ^{2} }$. To evaluate this integral, it is convenient to employ the standard property of the Dirac delta under a change of variables. For a continuously differentiable function $g(x)$ with simple roots $x_{i}$, one can write
\begin{align}
    \delta (g(x)) = \sum _{i} \frac{\delta (x-x_{i})}{\vert g' (x _{i}) \vert } . \label{identity_deltas}
\end{align}
This identity allows us to reduce the integration over $g(x)$ to a sum over the roots of the function, which considerably streamlines the calculation. Using the identity (\ref{identity_deltas}) we obtain
\begin{align}
    \delta \left( \tilde{\mu} - \sqrt{ \kappa ^{2}  + \lambda ^{2} } \right) = \frac{\tilde{\mu}}{\sqrt{ \tilde{\mu} ^{2}  - \lambda ^{2} - \kappa _{x} ^{2} }} \; \delta \left(  \kappa _{y}  - \sqrt{ \tilde{\mu} ^{2}  - \lambda ^{2} - \kappa _{x} ^{2} } \right)
\end{align}
Therefore, the the $\kappa _{y}$-integration reduces to a single contribution at the allowed root, provided that $\tilde{\mu} ^{2} > \lambda ^{2} + \kappa _{x} ^{2}$ (to guarantee that $\kappa_{y}$ is real valued), acoompanied by the prefactor $ \tilde{\mu} / \sqrt{ \tilde{\mu} ^{2}  - \lambda ^{2} - \kappa _{x} ^{2} }$, i.e.
\begin{align}
    G_{2} ^{\mbox{\scriptsize H}} (\mu ) =   \frac{  e ^{3}}{ \pi h  } \, \frac{ \lambda L_{y}}{\hbar v _{F}} \, \frac{\mbox{sgn}(\mu)}{\tilde{\mu} ^{2} } \,  \Theta ( \tilde{\mu} - \vert \lambda \vert ) \int _{0} ^{\sqrt{ \tilde{\mu} ^{2} - \lambda ^{2} } } d \kappa _{x}  \,  \frac{  [  \left( \tilde{\mu} / \kappa _{x}   \right) ^{2} - 1 ] \, \sin ^{2} \kappa _{x} }{  [ \cos ^{2} \kappa _{x} +  \left( \tilde{\mu} / \kappa _{x}   \right) ^{2} \, \sin ^{2} \kappa _{x}  ] ^{2} } \frac{1}{\sqrt{ \tilde{\mu} ^{2}  - \lambda ^{2} - \kappa _{x} ^{2} }} , \label{resulting_integral} 
\end{align}
The above expression is meaningful only when $\tilde{\mu} > \vert \lambda \vert $, as stated by teh Heaviside step function $\Theta (z)$, ensuring that the argument of the square root is positive and the integration range is finite. In contrast, if $\tilde{\mu} \leq \vert \lambda \vert $, the condition cannot be satisfied and the contribution vanishes, since there are no available states on the Fermi surface.

The resulting integral (\ref{resulting_integral}) cannot be expressed in terms of simple closed-form functions, which makes an analytic evaluation intractable. To facilitate a more reliable numerical treatment, it is convenient to manipulate the expression so as to remove the square-root singularity in the denominator. This can be achieved by introducing the substitution $\kappa _{x} = \sqrt{ \tilde{\mu} ^{2}  - \lambda ^{2} } \, \sin \theta $, with $\theta \in [0, \pi / 2]$. With this change of variables the expression for the conductance then reduces to
\begin{align}
    G_{2} ^{\mbox{\scriptsize H}} (\mu ) =   \frac{  e ^{3}}{ \pi h  } \, \frac{ \lambda L_{y}}{\hbar v _{F}} \, \frac{\mbox{sgn}(\mu)}{\tilde{\mu} ^{2} } \,  \Theta ( \tilde{\mu} - \vert \lambda \vert ) \int _{0} ^{ \pi / 2 } d \theta  \;  \frac{  [  \left( \tilde{\mu} / \Delta \right) ^{2} \csc ^{2} \theta - 1 ] \, \sin ^{2} ( \Delta \sin \theta ) }{  [ \cos ^{2} ( \Delta \sin \theta ) +  \left( \tilde{\mu} / \Delta \right) ^{2} \csc ^{2} \theta \, \sin ^{2} ( \Delta \sin \theta )  ] ^{2} }  , \label{final_Hall_conductance} 
\end{align}
where $\Delta = \sqrt{ \tilde{\mu} ^{2}  - \lambda ^{2} } $. This transformation not only eliminates the square-root singularity but also provides a numerically stable representation on a finite interval, ideally suited for further analysis and computation.

\begin{figure}
    \centering
    \includegraphics[width=0.5\linewidth]{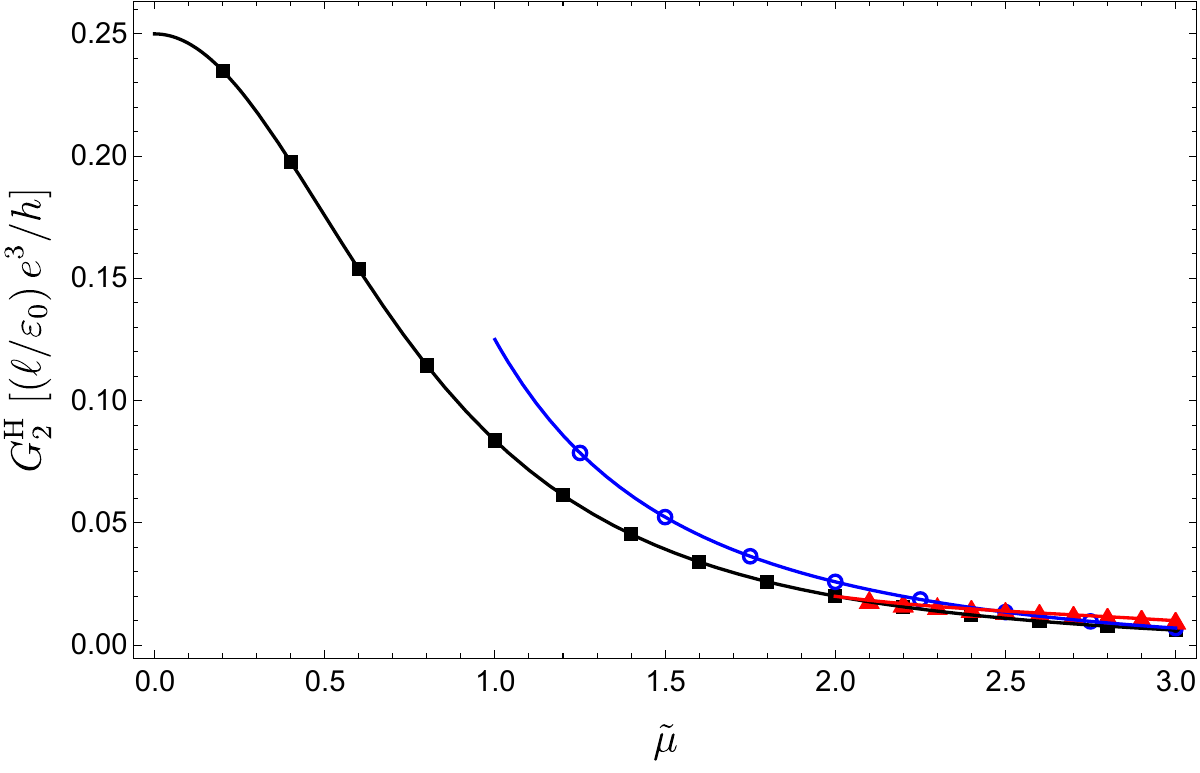}
    \caption{Second-order nonlinear Hall conductance $G_{2}^{\mathrm{H}}$ as a function of the rescaled chemical potential $\tilde{\mu}$. The gapless case ($\lambda=0$, black squares) shows a finite response that decays smoothly with doping, while the gapped phases ($\lambda=1$, blue circles; $\lambda=2$, red triangles) exhibit vanishing conductance inside the gap and a rapid onset at the band edges.}
    \label{nonlinear_Hall_conductances}
\end{figure}

Figure \ref{nonlinear_Hall_conductances} shows the second-order nonlinear Hall conductance $G_{2}^{\mathrm{H}}$ (in units of $(\ell/\varepsilon_{0}) e^{3}/h$) as a function of the rescaled chemical potential $\tilde{\mu}$. In the gapless case ($\lambda=0$, black squares), the response is finite and decays smoothly with increasing chemical potential, reflecting the continuous availability of carriers across the Dirac spectrum. By contrast, when a finite gap is present ($\lambda=1$, blue circles; $\lambda=2$, red triangles), the nonlinear Hall conductance vanishes inside the insulating regime and only becomes finite once the chemical potential enters the conduction band. { This behavior is fundamentally different from the linear Hall effect: while the quantized Hall conductance originates from a Fermi-sea property and remains constant throughout the gap, the nonlinear Hall response is governed by Fermi-surface quantities, such as the Berry-curvature dipole, and therefore requires accessible states at the chemical potential \cite{sodemann_quantum_2015,du_quantum_2021,kawabata_nonlinear_2022,MooreOrenstein2010} }. As a result, the gapped phases exhibit a clear suppression of the nonlinear Hall conductance within the gap, followed by a rapid onset at the band edge, consistent with the features displayed in the figure.

Although the integral expression in Eq. (\ref{final_Hall_conductance}) cannot be evaluated in closed form, a useful analytic information can be obtained in limiting regimes of the parameter $\Delta = \sqrt{ \tilde{\mu} ^{2}  - \lambda ^{2} } $. In the limit $\Delta \gg 1$, the phase $\sin ^{2} ( \Delta \sin \theta )$ oscillates rapidly, and the integrand can be approximated by replacing $\sin ^{2} ( \Delta \sin \theta )$ and $\cos ^{2} ( \Delta \sin \theta )$ with their average values. This procedure yields an analytic expression in which the oscillatory behavior is effectively smoothed out. The result shows that the nonlinear Hall conductance approaches
\begin{align}
    G_{2} ^{\mbox{\scriptsize H}} (\mu ) \approx  \frac{  e ^{3}}{   h  } \, \frac{ \lambda L_{y}}{\hbar v _{F}} \, \frac{\mbox{sgn}(\mu)}{\tilde{\mu} ^{2} } \,  \Theta ( \tilde{\mu} - \vert \lambda \vert ) \, \left[ \frac{ \tilde{\mu} (3 \tilde{\mu} ^{2} - 2 \lambda ^{2} ) }{(2 \tilde{\mu} ^{2} - \lambda ^{2} ) ^{3/2} } - 1 \right],
\end{align}
up to subleading corrections of order $\Delta ^{-1/2}$. This expression captures the large-$\Delta$ behavior, where the system exhibits a quasi-classical response dominated by the averaged Berry curvature and transmission.

In the opposite limit, $\Delta \ll 1$, the argument of the trigonometric functions is small, allowing for a systematic expansion. To leading order, one finds that the integrand simplifies to a smooth function of $\sin \theta$, and the resulting integral can be carried out analytically. The conductance then reduces to
\begin{align}
    G_{2} ^{\mbox{\scriptsize H}} (\mu ) \approx  \frac{  e ^{3}}{   h  } \, \frac{ \lambda L_{y}}{\hbar v _{F}} \, \mbox{sgn}(\mu) \,  \Theta ( \tilde{\mu} - \vert \lambda \vert ) \,  \frac{ \tilde{\mu} ^{2} + \lambda ^{2}   }{4 \tilde{\mu} ^{2} \, (1 + \tilde{\mu} ^{2} ) ^{2} } .
\end{align}
This expression shows that the nonlinear Hall conductance remains finite as $\Delta \to 0$, with its magnitude controlled by the balance between the effective mass parameter and the position of the chemical potential.

Taken together, these limiting forms provide analytic benchmarks that complement the full numerical evaluation of Eq. (\ref{final_Hall_conductance}). They also offer physical insight into how the nonlinear Hall response interpolates between the oscillatory, quasi-classical regime at large $\Delta$ and the smooth, perturbative regime at small $\Delta$.

The impact of dephasing and structural disorder on the nonlinear Hall conductance will be analyzed in Sec.~VI, where we demonstrate that the vanishing response inside the gap and the edge-onset behavior are qualitatively preserved.

\section{Effects of dephasing and disorder}

In realistic devices, electronic transport is inevitably affected by sources of dephsing and structural disorder. To account for this, we now analyze how these effects modify the transmission function and the resulting conductance.  Dephasing arises from inelastic scattering with phonons, magnons, or other low-energy excitations, as well as from slow electrostatic fluctuations of the local potential. Such processes randomize the additional phase acquired by carriers as they traverse the topological slab, thereby suppressing the Fabry-Pérot oscillations that appear in the fully coherent case discussed above. Structural disorder, such as interface roughness or local variations in the barrier potential, produces similar effects by intrducing sample-to-sample fluctuations of the effective propagation length or barrier height. Experimentally, these mechanisms are characterized by the phase-coherence length $\ell _{\phi}$, which sets the typical distance over which quantum coherence is preserved \cite{datta_electronic_1995,Imry2002}.

A convenient way to incorporate these effects into the Landauer framework is through statistical averaging over random phase shifts. If the oscillatory factor $\cos ( 2 q _{x} L)$ is replaced by $\cos (2 q _{x} L + \delta \phi )$, with $\delta \phi$ drawn from a Gaussian distribution of variance $\langle \delta \phi ^{2} \rangle = 2L / \ell _{\phi} $, the average yields 
\begin{align}
    \langle \cos (2 q _{x} L + \delta \phi ) \rangle 
    = e ^{-2L/\ell _{\phi}} \cos (2 q _{x} L ) , 
    \label{eq:phase_avg}
\end{align}
which interpolates smoothly between the coherent limit ($\ell _{\phi} \to \infty$) and the fully incoherent regime ($\ell _{\phi} \ll L$), where oscillations vanish altogether \cite{Beenakker1997,Bardarson2007}. Additional static disorder sources, such as barrier-width fluctuations $\delta L$ or gate-potential fluctuations $\delta V$, can be incorporated in the same spirit by multiplying Eq.~(\ref{eq:phase_avg}) with Gaussian damping factors $\exp[- ( q _{x} \sigma _{L}) ^{2} ]$ or $\exp \{-L ^{2} (\partial q _{x} / \partial V ) ^{2} \sigma _{V} ^{2} \}$, respectively.

The transmission probability (\ref{Transmission}) modified by dephasing then reads
\begin{align}
    T _{\phi} (E,k_{y}) = \frac{ 1 }{  C _{\phi} ^{2} (q _{x} L) +  \left( \frac{E}{ \hbar v _{F}  q _{x} }  \right) ^{2} \, S _{\phi} ^{2} (q _{x} L) } ,  \label{Transmission_dephasing}
\end{align}
where we introduced the averaged functions
\begin{align}
    C _{\phi} ^{2} ( x ) = \tfrac{1}{2} \Big[ 1 + e ^{-2L / \ell _{\phi}} \cos ( 2 x ) \Big] , \qquad S _{\phi} ^{2} ( x ) = \tfrac{1}{2} \Big[ 1 - e ^{-2L / \ell _{\phi}} \cos ( 2 x ) \Big] .
\end{align}
This phenomenological damping scheme has been widely used in mesoscopic transport to mimic the effect of decoherence \cite{datta_electronic_1995,Imry2002,Bardarson2007}, and provides a simple yet physically transparent way to assess the robustness of our predictions against imperfections.

Importantly, the dephasing factor acts only on the oscillatory phase of propagating modes ($q _{x} \in \mathbb{R}$). In the evanescent sector ($q _{x} = i \kappa$), the longitudinal dynamics is hyperbolic ($\cos \to \cosh$, $\sin \to \sinh$) and no phase averaging is applied; the corresponding transmission is obtained by analytic continuation of Eq.~(\ref{Transmission}) and is dominated by the exponential decay $\sim e^{-2\kappa L}$ rather than by interference. This phenomenological damping scheme has been widely used in mesoscopic transport to mimic the effect of decoherence~\cite{datta_electronic_1995,Imry2002,Bardarson2007}, and provides a simple yet physically transparent way to assess the robustness of our predictions against imperfections.

To make contact with observables, in what follows we focus on the propagating contribution to the longitudinal conductance, where dephasing induces quantitative changes. Specifically, we will present (i) the linear conductance $G_{1 \phi}^{(p)}(\tilde{\mu})$ and (ii) the nonlinear conductances $G_{2 \phi}^{(p)}(\tilde{\mu})$ and $G _{3 \phi}^{(p)}(\tilde{\mu})$ computed from Eq.~(\ref{Transmission_dephasing}), and compare them with their fully coherent counterparts to highlight the progressive suppression of Fabry-Pérot oscillations while preserving the overall trends. 
(Results for the evanescent contribution are unchanged by phase averaging and are therefore omitted for brevity.)

\begin{figure}
    \centering
    \includegraphics[width=0.45\linewidth]{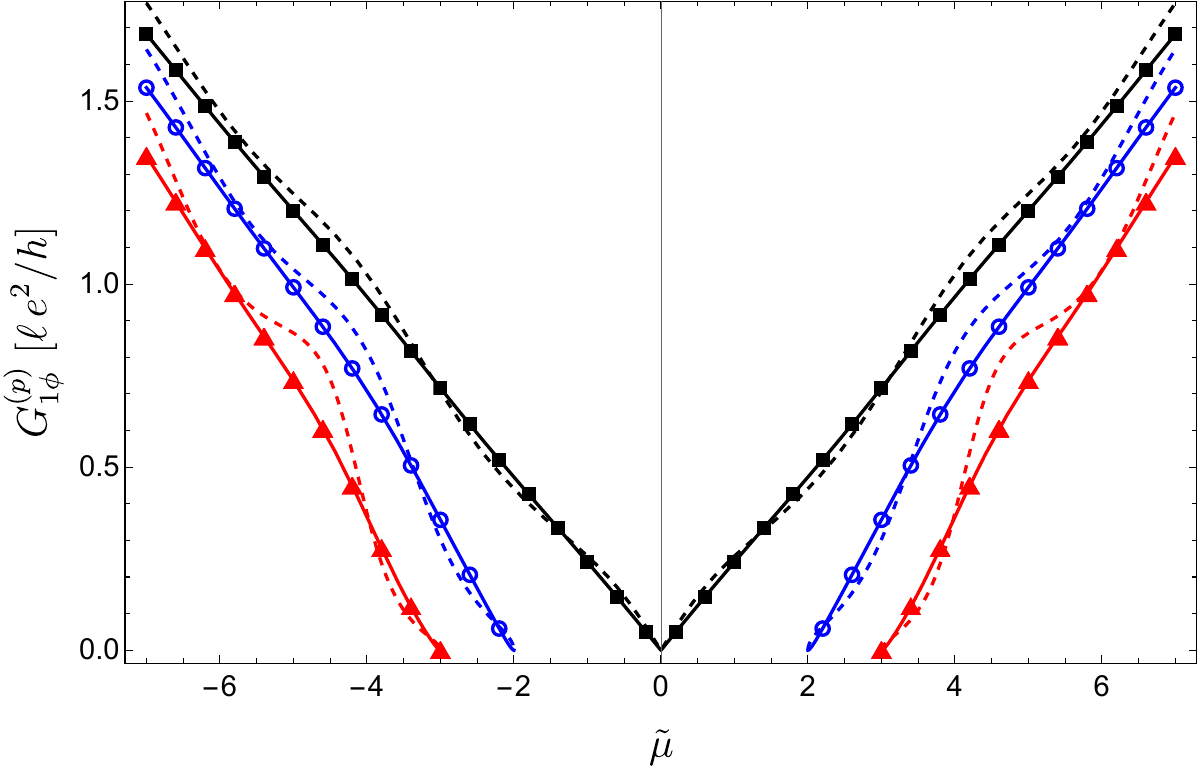} \hspace{1cm}
    \includegraphics[width=0.45\linewidth]{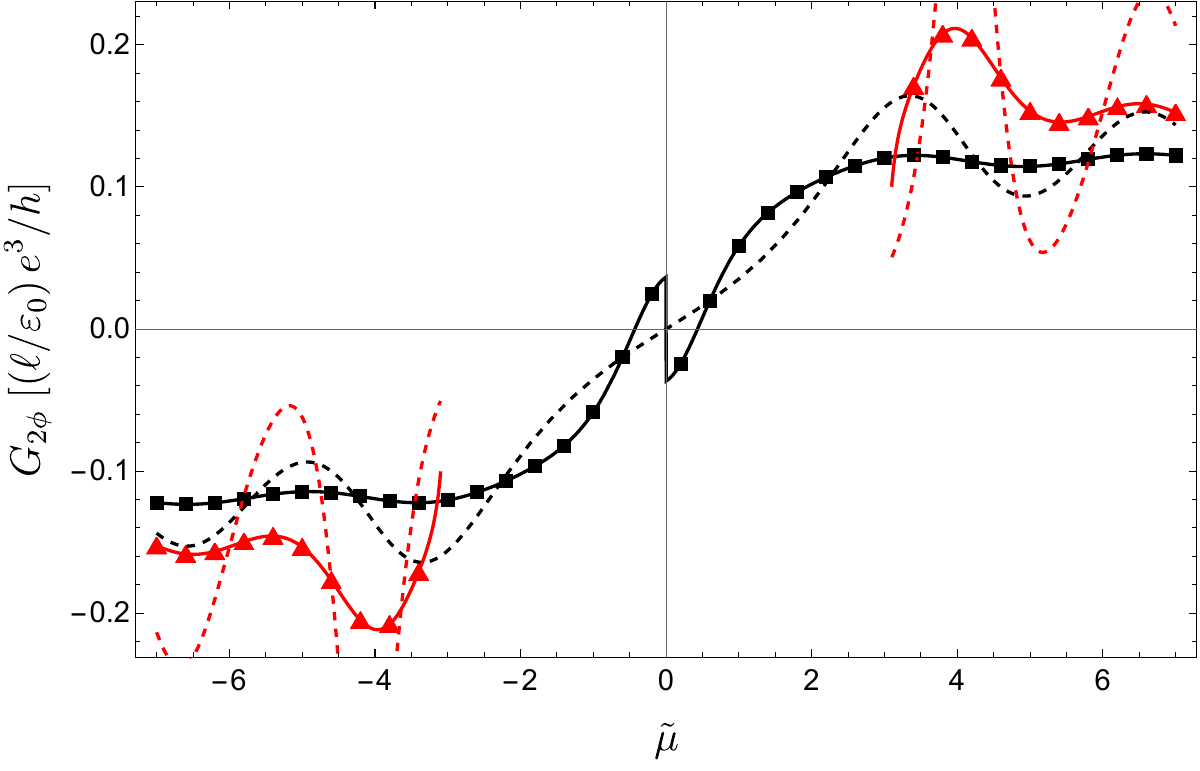}\\
    \includegraphics[width=0.45\linewidth]{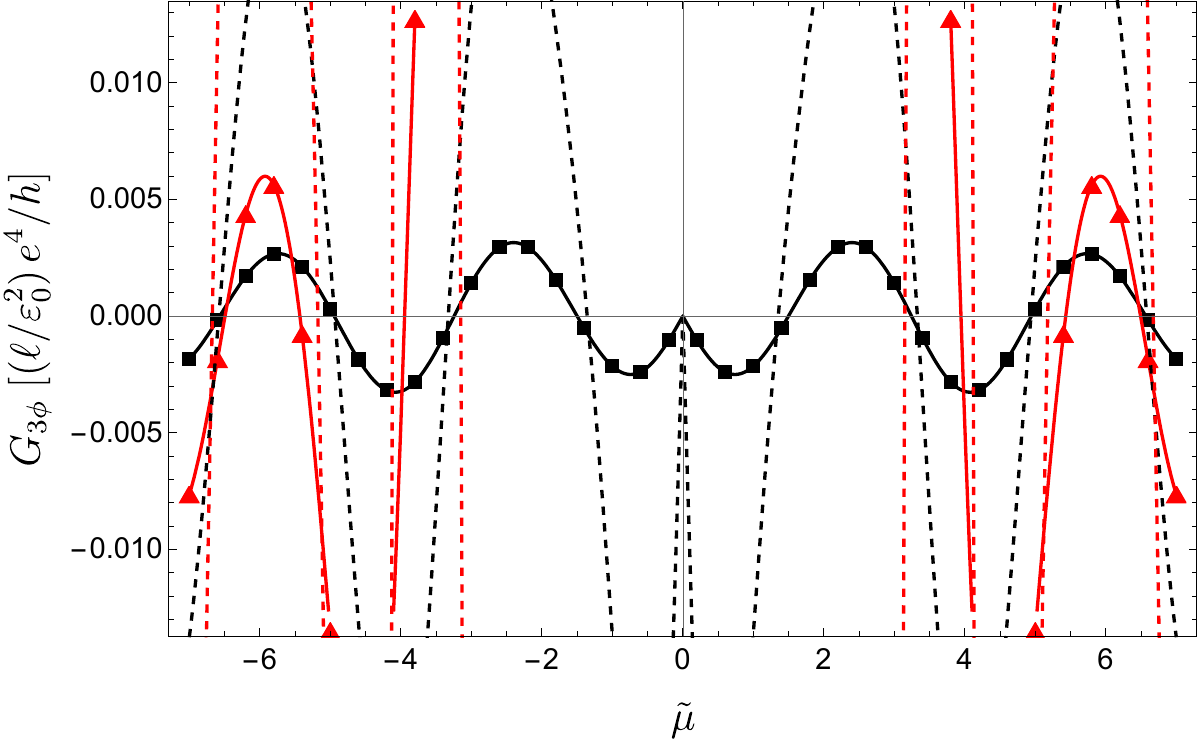}
    \caption{Propagating contribution to the longitudinal conductance with and without dephasing. 
Top left: linear $G _{1\phi}^{(p)}(\tilde{\mu})$ with $\lambda = 0$ (black), $\lambda=2$ (blue), and $\lambda=3$ (red). 
Top right: second-order $G_{2\phi}^{(p)}(\tilde{\mu})$.  Bottom: third-order $G_{3\phi}^{(p)}(\tilde{\mu})$. 
In the nonlinear panels, only the gapless case ($\lambda=0$, black) and one representative massive case ($\lambda=3$, colored) are shown to keep the plots readable and emphasize dephasing effects. 
Solid lines include dephasing with $\ell_{\phi}=1$, whereas dashed lines represent the coherent limit ($\ell _{\phi} \to \infty$). 
Dephasing primarily damps Fabry-Pérot oscillations while leaving the overall doping trends intact.} \label{longitudinal_conductance_dephasing}
\end{figure}

Figure~\ref{longitudinal_conductance_dephasing} shows the propagating contribution to the linear longitudinal conductance $G _{1\phi}^{(p)}(\tilde{\mu})$ as a function of the rescaled chemical potential.  The black curves correspond to the gapless case ($\lambda = 0$), representative of graphene, while the blue and red curves correspond to Chern insulating phases with finite mass parameters $\lambda = 2$ and $\lambda = 3$, respectively. In each case, the solid lines include dephasing with a phase-coherence length $\ell _{\phi} = 1$, whereas the dashed lines display the fully coherent limit ($\ell _{\phi} \to \infty$). The comparison highlights two key trends. First, the opening of a mass gap systematically reduces the overall  conductance and shifts the onset of transmission to larger values of $|\tilde{\mu}|$, reflecting the suppression of propagating channels near charge neutrality. Second, dephasing produces a moderate reduction of the conductance across the entire doping range, more visible at intermediate values of $\tilde{\mu}$ where Fabry-Pérot oscillations contribute significantly. At large doping, where transport is dominated by extended states, the difference between the coherent and dephased curves becomes less pronounced. This behavior illustrates that phase coherence primarily enhances the interference features, while the gross magnitude of the conductance remains robust. For clarity, in the nonlinear conductance plots presented below we restrict the analysis to two representative cases (gapless $\lambda = 0$ and one finite mass $\lambda = 2$), which makes the dephasing-induced changes easier to visualize.

\begin{figure}
    \centering
    \includegraphics[width=0.45\linewidth]{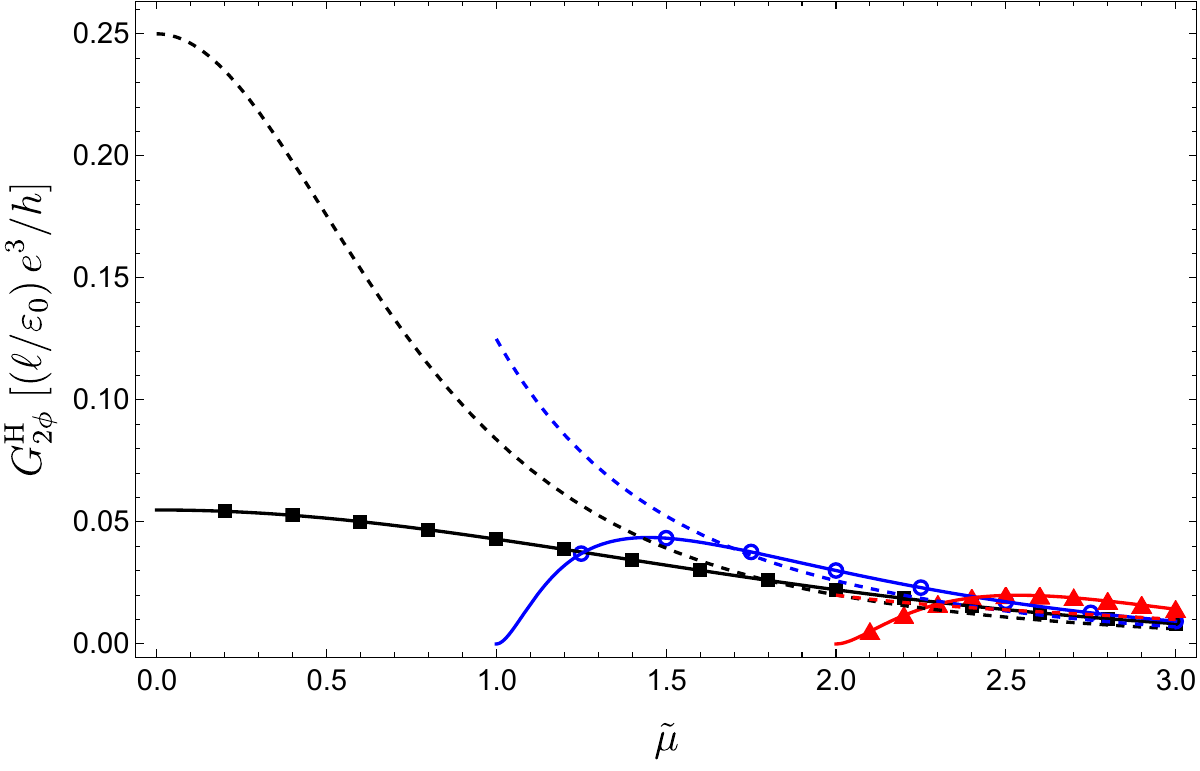} 
    \caption{Nonlinear Hall conductance $G^{\mathrm{H}}_{2\phi}(\tilde{\mu})$ with (solid)  and without (dashed) dephasing. Dephasing reduces the overall magnitude of the response, most visibly near the band edge, while the qualitative trends remain unchanged.} \label{Hall_conductance_dephasing}
\end{figure}

Figure~\ref{Hall_conductance_dephasing} shows the nonlinear Hall conductance $G^{\mathrm{H}}_{2\phi}(\tilde{\mu})$ for different values of the mass parameter. Dashed lines correspond to the fully coherent case ($\ell_{\phi}\to\infty$), while solid lines include dephasing with $\ell_{\phi}=1$. The comparison reveals that dephasing systematically suppresses the magnitude of the nonlinear Hall response across the doping range. In the gapless case ($\lambda=0$, black curves), the suppression is most visible at small chemical potential, where the coherent signal reaches its maximum. For the gapped phases ($\lambda=1$ blue, $\lambda=2$ red), the effect of dephasing is to lower the amplitude of the peaks that emerge once the chemical potential exceeds the band edge, thereby reducing the overall nonlinear response. At larger $\tilde{\mu}$, where the conductance already decays towards zero due to the diminishing role of Berry curvature hot spots, the difference  between the coherent and dephased curves becomes negligible. These results highlight that phase coherence enhances the nonlinear Hall signal, while decoherence acts as a damping mechanism that  smooths out its most pronounced features.

It is important to emphasize that the suppression of the nonlinear Hall conductance by dephasing does not imply a breakdown of the underlying topological protection. In the linear regime, the quantized Hall conductance inside the gap is guaranteed by the Chern number and remains robust against disorder and decoherence, as long as the bulk gap is not closed. In contrast, the nonlinear Hall response does not originate from a quantized topological invariant but from geometric properties of the Bloch bands, such as the Berry  curvature distribution and its associated dipole moment. As a consequence, the magnitude of $G^{\mathrm{H}}_{2}$ is sensitive to phase coherence and other sample-specific details, and dephasing acts as a damping mechanism that reduces its amplitude. The comparison between the linear and nonlinear cases thus illustrates a fundamental distinction: while the linear Hall response is topologically protected and therefore universal, the nonlinear Hall response is geometric in nature and consequently more fragile, though it still reflects the underlying Berry curvature structure of the Chern phase.

\section{Summary and conclusions} \label{Conclusions}

Topological phases of matter have emerged as a central theme in modern condensed matter physics, both for their fundamental significance and for their potential applications in electronic devices \cite{hasan_colloquium_2010, qi_topological_2011}. Among them, Chern insulators realize quantized Hall transport without an external magnetic field, making them ideal platforms for exploring novel transport regimes \cite{haldane_model_1988, yu_quantized_2010, chang_experimental_2013}. In this work, we studied a trivial-topological-trivial junction described by the continuous Qi-Wu-Zhang model \cite{qi_topological_2006}, where a topological Chern-insulating slab is coupled to trivial leads and subjected to a tunable electrostatic barrier. Within the Landauer framework we obtained the full transmission amplitude by solving directly the Dirac-like equation, and demonstrated that this system exhibits Klein tunnelling despite the presence of a bulk energy gap. This phenomenon, originally associated with relativistic fermions \cite{DOMBEY199941, katsnelson_chiral_2006}, here arises from the spinor matching enabled by band inversion across the junction, encoded in the sign reversal of the effective Dirac mass in the central region. Within the Qi-Wu-Zhang model, this mass inversion corresponds to the transition between trivial and Chern-insulating phases, providing a direct connection between topological band inversion and the resulting scattering properties. {In this sense, the perfect and near-perfect transmission found here reflects the band-inverted nature of the junction and its associated Dirac spinor structure, rather than relying on gapless bulk states.}

Using the transmission function as input, we systematically analyzed the longitudinal conductance. The linear response is entirely determined by the total transmission, while the nonlinear conductances involve successive energy derivatives of the transmission evaluated at the chemical potential. We obtained both numerical and approximate analytic expressions for these quantities, valid in the asymptotic regimes of large chemical potential and near the band edge. These results establish clear benchmarks for identifying the crossover between quasi-classical oscillatory transport and smooth perturbative behavior. Importantly, our analysis highlights how barrier height, slab thickness, and the effective mass parameter can be tuned to achieve strong nonlinear responses and optimal rectification ratios. These findings demonstrate how band inversion, encoded in the mass profile of the heterostructure, governs the transmission properties and provides a controllable route to engineering nonlinear transport in Chern-insulator junctions.

We further extended the formalism to include the transverse Hall conductance, making explicit the role of the anomalous velocity and Berry curvature in driving nonlinear Hall effects within the Landauer framework. The derived expressions for the quadratic and higher-order Hall conductances demonstrate that the nonlinear response vanishes both in the ballistic limit and in the tunneling regime, attaining maximum values at intermediate transmission probabilities. Analytical approximations in the limits of large and small chemical potential once again provided physical insight, complementing the numerical evaluation. Together, these results give a comprehensive picture of linear and nonlinear transport in topological heterostructures. {In particular, the Hall response depends explicitly on the Berry curvature of the Chern-insulating region, providing a direct transport manifestation of the underlying topological band structure. This establishes a clear distinction between the longitudinal transmission, governed by band inversion and spinor matching, and the transverse nonlinear response, which directly probes Berry-curvature physics.}

Finally, we connected our theoretical findings to realistic material platforms, specifically magnetic topological insulators based on Cr-doped (Bi,Sb)$_2$Te$_3$, where the central region realizes a Chern-insulating phase ($C= \pm 1$) while the leads remain trivial ($C=0$). This correspondence suggests that the phenomena discussed here are within reach of current experimental techniques, particularly in thin-film geometries where gate voltages can be used to tune the electrostatic barrier. Looking forward, it will be of interest to examine the robustness of these results against disorder and electron-electron interactions, as well as to extend the analysis to time-dependent driving fields. Experimental observation of Klein tunnelling and nonlinear Hall transport in such heterostructures would represent a significant advance, confirming the predictions made here and paving the way toward exploiting topological materials in future electronic and spintronic devices. {More broadly, our results establish Chern-insulator heterostructures as a versatile platform for investigating how band inversion, Berry curvature, and quantum-coherent scattering combine to shape linear and nonlinear transport, opening new directions for both fundamental studies and device-oriented applications.}

\acknowledgements{A.M.-R. acknowledges financial support by UNAM-PAPIIT project No. IG100224, UNAM-PAPIME project No. PE109226, by SECIHTI project No. CBF-2025-I-1862 and by the Marcos Moshinsky Foundation.}

\bibliography{References.bib}
\end{document}